\UseRawInputEncoding
\documentclass[superscriptaddress, notitlepage, reprint,twocolumn]{revtex4-1}
\usepackage[english]{babel}
\usepackage{amsmath,amsthm,amssymb,lipsum}
\usepackage{amsfonts}
\usepackage[pdfborder={0 0 0}, colorlinks=true, urlcolor=blue, linkcolor=blue, citecolor=blue]{hyperref}
\usepackage{color}
\usepackage{graphicx}
\usepackage{float}
\usepackage{subfigure}
\usepackage{lipsum}
\usepackage{epstopdf}
\usepackage{dcolumn}
\usepackage{mathrsfs}
\usepackage{amsmath,amssymb}

\begin{document}

\title{Quantum sensing of a weak drive signal in cascaded quantum van der Pol oscillators}
\author{Dong  Xie}
\email{xiedong@mail.ustc.edu.cn}
\author{Chunling Xu}
\affiliation{College of Science, Guilin University of Aerospace Technology, Guilin, Guangxi 541004, People's Republic of China}

\begin{abstract}
We have constructed a quantum sensing system composed of two quantum van der Pol (vdP) oscillators. Coherent and dissipative couplings are combined to establish a generic nonreciprocal coupling between the two vdP subsystems. A weak driving signal to be measured is applied to an oscillator (probe system) and then transferred via non-reciprocal coupling to another oscillator (measurement system) that is easier to measure. This allows for the acquisition of information about the driving strength of the signal being measured. For the classic cascaded dual vdP oscillators, complete non-reciprocal coupling yields a much higher susceptibility than a single classical vdP oscillator alone. For the quantum vdP system, the sensitivity generally does not diverge at the zero driving force point. When only dissipative coupling and two-particle loss are present, the system will have a dark mode at the zero driving force point, which enables the measurement precision to ultimately reach the Heisenberg limit in the time domain. Enhancing incoherent driving can effectively resist dissipation and thereby improve measurement precision. Moreover, in the strong incoherent driving regime, the improvement factor of non-reciprocal coupling over reciprocal coupling in terms of measurement precision tends to infinity as the two-particle loss rate of the measurement system approaches zero.

\end{abstract}

\maketitle
\section{Introduction}
Quantum sensing constitutes a prominent research frontier at the interface of quantum science and precision metrology. Its goal is to convert coherence, squeezing, entanglement, and quantum manipulation into high-precision estimations of weak physical quantities under the constraints of quantum noise, measurement feedback, and resource limitations. In recent years, quantum sensing has not only advanced fundamental physical tests but also provided key technical support for the development of quantum communication, quantum computing, and low-power quantum devices \cite{lab1,lab2,lab3,lab4,lab5,lab6,lab7,lab8,lab9}. Among various physical carriers, resonators have become an important platform for precise measurement and weak signal detection due to their adjustable frequency, high quality factor, and good compatibility with superconducting circuits, optical mechanical systems, and cavity quantum electrodynamics platforms \cite{lab9,lab10}. Therefore, how to measure a weak driving signal that is much smaller than the thermal noise or detection noise level is one of the core issues in resonator quantum sensing research.

Realistic quantum systems inevitably couple to thermal reservoirs, measurement channels and control lines, giving rise to ubiquitous dissipation. Most studies adopt linear damping and Markov approximation, using the Lindblad master equation to describe the energy exchange of the system with the environment \cite{lab11}. However, many practical resonators exhibit amplitude-dependent damping: the system gains energy at small amplitudes and is suppressed at large amplitudes due to nonlinear mechanisms, eventually forming stable self-sustained oscillations. This behavior is typically described by the vdP model, whose classical form includes the competition between negative linear damping and cubic nonlinear damping \cite{lab12}. When approaching the self-sustained oscillation threshold or the low amplitude limit, the response of the classical vdP oscillator can become very steep, and weak drive can even lead to sensitivity divergence. This nonlinear amplification mechanism is believed to be related to the high-sensitivity active amplification of biological mechanical receptors such as cochlear hair bundle cells, enabling biological systems to detect extremely weak mechanical stimuli in strong background noise \cite{lab13,lab14}.
After quantizing the vdP oscillator, zero-point fluctuations, transitions to the heat reservoir, quantum measurement feedback, and finite excitation numbers introduce noise that is absent in the classical model. The steady state of the quantum vdP oscillator is no longer completely described by a single deterministic trajectory; its amplitude diffusion, phase diffusion, and synchronization behaviors need to be studied using quantum master equations and phase space distributions, etc. Existing theoretical studies have analyzed the locking and synchronization thresholds of quantum vdP oscillators using quantities such as quantum phase space distributions and two-oscillator correlations \cite{lab15,lab16}. These results indicate that the zero-drive sensitivity divergence observed in the classical model cannot be directly extrapolated to the quantum regime: quantum noise changes the response scale near the threshold and creates a new trade-off between ultra-high sensitivity and measurement stability.
When the investigation extends from a single oscillator to two coupled oscillators, information transfer pathways and noise backflow emerge as critical factors governing sensing performance. Non-reciprocal coupling can establish directional information flow between coupled resonators or cavity systems, allowing signals to be transmitted from one oscillator to another while suppressing disturbances and measurement feedback in the opposite direction. This provides the possibility of constructing directional weak signal sensors and unidirectional quantum transducers \cite{lab17,lab18}. Meanwhile, studies of two-oscillator quantum synchronization \cite{lab19,lab20,lab21,lab22} focus on whether two self-sustaining oscillators can establish a stable phase relationship and how this phase relationship is affected by quantum noise and dissipation. These two lines of research respectively indicate that non-reciprocal coupling facilitates directional information transmission and that the quantum vdP oscillator serves as an important platform for studying quantum synchronization. One key question arising from this is: When two quantum vdP oscillators are connected through non-reciprocal coupling, can the non-reciprocal properties improve the quantum sensing precision of weak driving signals?

In this article, we employ two vdP oscillators to implement quantum sensing measurements. A weak coherent drive acts on one vdP resonator. Through coherent and dissipative coupling, the information of the drive is transferred to the other resonator. The amplitude information of the drive is then extracted via homodyne measurement. When the incoherent drive and single-particle loss are both smaller than the two-particle loss, the measurement precision is limited and will not diverge. Complete non-reciprocal coupling will result in a measurement precision that is twice as high as that obtained with reciprocal coupling.  When the incoherent drive and dissipation, as well as the coherent coupling strength are all equal to 0, the dark mode forms under zero coherent drive. This is beneficial for the measurement of extremely weak drive, ultimately achieving the Heisenberg limit of measurement precision in the time domain. Under strong incoherent drive, the measurement precision increases as the two-particle loss rate decreases, and the factor by which the non-reciprocal coupling exceeds the reciprocal coupling also increases as the two-particle loss rate decreases. At the critical incoherent driving condition, only when the weak driving force is near zero, can the measurement uncertainty obtained with nonreciprocal coupling tend to 0 as the two-particle loss rate approaches 0. At this point, compared with reciprocal coupling, the improvement in measurement precision obtained through non-reciprocal coupling will also increase as the two-particle loss rate decreases.

The rest of this article is organized as follows. In section II, the combination of coherent and dissipative couplings realizes the nonreciprocal coupling in the cascaded vdP oscillators. In section III, we obtain finite measurement precision by employing weak incoherent drive for the quantum vdP resonator. When only dissipation coupling is present, a dark mode appears at zero drive in section IV. In section V, measurement precision is obtained under strong incoherent driving conditions. In section VI, the sensing precision for weak drives at critical incoherent drive is investigated. We make a conclusion in section VII.
\section{nonreciprocal coupling in vdP oscillators}
We consider two resonant quantum vdP oscillators coupled together via coherent and dissipative couplings, whose corresponding dynamical equations can be written in the form of a Lindblad master equation under the Markov approximation. In the rotating frame, the density matrix is described by \cite{lab18,lab23}
\begin{align}
\dot{\rho}=&-i[H_d+H_c,\rho]+\gamma_{1}^+\mathcal{D}_{a^\dagger}[\rho]+\gamma_{1}^-\mathcal{D}_a[\rho]+\gamma_{2}\mathcal{D}_{a^2}[\rho]\nonumber\\
&+\Gamma_{1}^+\mathcal{D}_{b^\dagger}[\rho]+\Gamma_{1}^-\mathcal{D}_b[\rho]+\Gamma_{2}\mathcal{D}_{b^2}[\rho]+G\mathcal{D}_{a+b}[\rho],
\end{align}
where $a$ and $b$ denote the annihilation operators of the two subsystems, the driven Hamiltonian $H_d=i\Omega(a^\dagger -a)$ with the driving strength $\Omega$ to be estimated, the superoperator $\mathcal{D}_c[\rho]=2c\rho c^\dagger-c^\dagger c\rho-\rho c^\dagger c$, $\gamma_{1}^+$ ($\Gamma_{1}^+$) denotes the one-particle gain rate, $\gamma_{1}^-$ ($\Gamma_{1}^-$) denotes the one-particle loss rate, $\gamma_{2}$ ($\Gamma_{2}$) denotes the two-particle loss rate, and $G$ denotes the dissipation coupling strength arising from the coupling to the shared reservoir.
The Hamiltonian of coherent coupling $H_c$ is given by
\begin{align}
H_c=ig(a^\dagger b-a b^\dagger)
\label{eq:2}
\tag{2}
\end{align}
where  $g$ denotes the coherent coupling strength.

According to the Lindblad master equation, the expectation vlaues of the operators $\langle a\rangle$ and $\langle b\rangle$ satisfy
\begin{align}
\langle\dot{a}\rangle&=(\gamma_{1}^+- \gamma_{1}^--G)\langle a\rangle +(g-G)\langle b\rangle-2\gamma_{a2}\langle  a^\dagger a^2\rangle+\Omega,\tag{3}\\
\langle\dot{b}\rangle&=(\Gamma_{1}^+- \Gamma_{1}^--G)\langle b\rangle -(g+G)\langle a\rangle-2\gamma_{b2}\langle  b^\dagger b^2\rangle.
\label{eq:4}
\tag{4}
\end{align}
When both dissipative and coherent couplings are present simultaneously, the effective coupling between the two systems becomes non-reciprocal, i.e., $g-G \neq g+G$.  The completely non-reciprocal coupling appears at $G=g$. At this point, the effective coupling becomes directional, meaning that the probe system $a$ influences the measurement system $b$, while $b$ has no effect whatsoever on $a$.

Replacing the operator $a\ (b)$ with the complex number $\alpha$  ($\beta$), the classical vdP oscillators are recovered.
When $\gamma_{1}^+- \gamma_{1}^--G=0$ and $\Gamma_{1}^+- \Gamma_{1}^--G=0$, the real solution for $\beta$ can be obtained in the limit $\Omega\rightarrow0$

\begin{align}
\beta&=\frac{g}{\Gamma_{2}}(\frac{\Omega}{2\gamma_{2}})^{1/9}, G=g;\tag{5}\\
\beta&=\frac{\Omega}{G-g}, |G-g|\gg\Omega.
\label{eq:6}
\tag{6}
\end{align}

The susceptibility $\chi_c=\frac{d\alpha}{d\Omega}$ can be further derived
\begin{align}
\chi_c&=-\frac{g}{9\Gamma_{2}}(\frac{1}{2\gamma_{2}\Omega^8})^{1/9}, G=g;\tag{7}\\
\chi_c&=\frac{1}{G-g},\ |G-g|\gg\Omega.\tag{8}
\end{align}

It shows that the susceptibility is divergent at zero drive $\chi_c\propto \Omega^{-8/9}$  with the completely non-reciprocal coupling. It surpasses the sensing precision previously achieved for a single harmonic oscillator ($\chi^s_c\propto \Omega^{-2/3}$) in Ref.~\cite{lab24}. This indicates that complete non-reciprocal coupling can significantly enhance the measurement susceptibility of weak driving signals in classical vdP oscillators.

\section{quantum vdP oscillators with weak incoherent driving }
In this section, we investigate whether the quantum vdP oscillator still exhibits divergent sensitivity at the zero coherent drive as $\Omega\rightarrow0$.

\begin{figure}[h]
\includegraphics[scale=0.40]{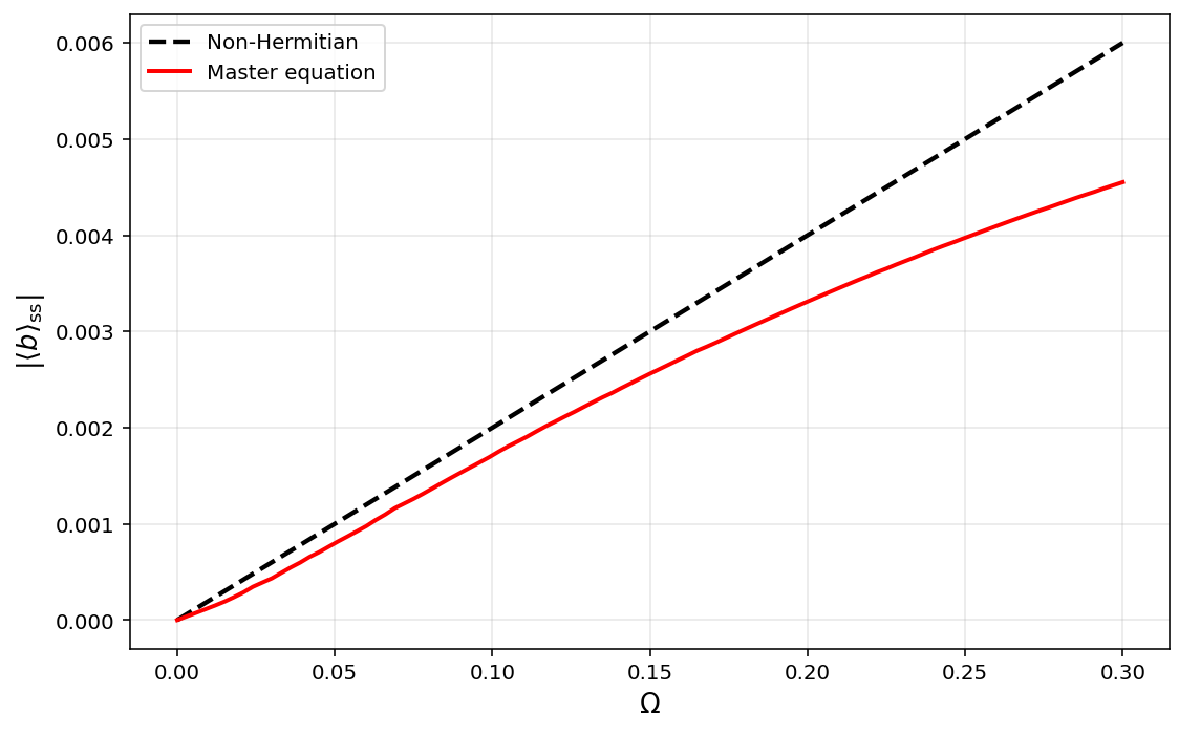}
 \caption{\label{fig.1} The comparison graph of the $b$-mode steady-state expectation value is obtained by using the method of equivalent non-Hermitian Hamiltonian and numerical solution of the Lindblad master equation. Here, the dimensionless parameters are chosen as: $\gamma_1^+=\Gamma_1^+=0.25, \gamma_1^-=\Gamma_1^-=0.05, \gamma_2=\Gamma_2=5.0, g=G=0.02$. }
\end{figure}

When the two-particle loss rates are large enough, i.e., $\gamma_{1}^+\ll\gamma_2$ and $\Gamma_{1}^+\ll\Gamma_2$, the dynamical evolution can be truncated to the lowest three energy levels of each subsystem.
In the weakly driven regime, that is, when $\Omega\ll1$, by using a non-Hermitian Hamiltonian\cite{lab25,lab26} and truncating to the lowest three energy levels for each subsystem, the expectation value of $b$ can be obtained. As shown in Fig.~\ref{fig.1}, the analytical results obtained via the non-Hermitian method and the numerical results from the master equation become increasingly consistent as $\Omega$ decreases.
At lowest order \(O(\Omega)\), the expectation values are (see the detail in Appendix A)
\begin{align}
\langle a\rangle
&=\frac{\beta\Omega}{\Delta_1}+O(\Omega^3),\tag{9}\\
\langle b\rangle&=-\frac{(g+G)\Omega}{\Delta_1}+O(\Omega^3),\tag{10}\\
\langle b^\dagger b\rangle&=-\frac{(g+G)^2\Omega^2}{g^4}+O(\Omega^3),\tag{11}
\end{align}
where $\Delta_1=\mu \nu+g^2-G^2,$ with $\mu=\gamma_{1}^++\gamma_{1}^-+G $ and $\nu=\Gamma_{1}^++\Gamma_{1}^-+G$.
Clearly, from the above  expressions, the quantum counterpart shows no divergent sensitivity at zero drive; instead, a finite value is obtained.

When $g^2+(\gamma_{1}^++\gamma_{1}^-)(\Gamma_{1}^++\Gamma_{1}^-)\gg g(\gamma_{1}^++\gamma_{1}^-+\Gamma_{1}^++\Gamma_{1}^-)$, the ratio of the susceptibility obtained through completely non-reciprocal coupling to that obtained through reciprocal coupling can reach twice, i.e., $\chi_{G=g}/\chi_{G=0}=2$. Due to the weak driving strength, the variance of the quadrature operator $M=(b^\dagger+b)$ is given by
\begin{equation}
\Delta M=\langle(b^\dagger+b)^2\rangle-\langle(b^\dagger+b)\rangle^2=1+O(\Omega^2).\tag{12}
\end{equation}
Then, we obtain the same sensitivity factor as that of the susceptibility
\begin{equation}
\delta \Omega_{G=g}/\delta \Omega_{G=0}=2.\tag{13}
\end{equation}

\section{Steady-state dark-mode approximation }

When $\gamma^+_1=\gamma^-_1=0$, $\Gamma^+_1=\Gamma^-_1=0$ and $g=0$,
Assume that $G>0$ is fixed. The bright mode $c=\frac{a+b}{\sqrt{2}}$
is subject to collective dissipation, whereas the dark mode $d=\frac{a-b}{\sqrt{2}}$
is not directly damped by the linear collective dissipator. Consequently, the steady state approaches $\beta\simeq-\alpha.$

We obtain the first-moment equations
\begin{align}
\frac{d}{dt}\langle a\rangle
&= \Omega - G\langle a+b\rangle - 2\gamma_2\langle a^\dagger a^2\rangle, \tag{14}\\
\frac{d}{dt}\langle b\rangle
&= -G\langle a+b\rangle - 2\gamma_2\langle b^\dagger b^2\rangle.\tag{15}
\end{align}

Under the factorization approximation
\begin{equation}
\langle a^\dagger a^2\rangle \simeq |\alpha|^2\alpha,
\qquad
\langle b^\dagger b^2\rangle \simeq |\beta|^2\beta,\tag{16}
\end{equation}
the semiclassical equations become
\begin{align}
\dot{\alpha}
&= \Omega - G(\alpha+\beta) - 2\gamma_2|\alpha|^2\alpha, \tag{17}\\
\dot{\beta}
&= -G(\alpha+\beta) - 2\gamma_2|\beta|^2\beta.\tag{18}
\end{align}
Assuming the steady-state amplitudes are real, we set
\begin{equation}
\alpha = s+q,
\qquad
\beta = -s+q,\tag{19}
\end{equation}
where \(s\) is the dark-mode amplitude and \(q\) is a small bright-mode component. Then
\begin{equation}
\alpha+\beta = 2q.\tag{20}
\end{equation}
The steady-state equations are
\begin{align}
& \Omega - 2Gq - 2\gamma_2(s+q)^3=0, \tag{21}\\
& 2Gq + 2\gamma_2(-s+q)^3=0.\tag{22}
\end{align}
Subtracting the two equations gives
\begin{equation}
 \Omega - 4\gamma_2 s^3 - 12\gamma_2 s q^2=0.\tag{23}
\end{equation}
Adding the two equations gives
\begin{equation}
 \Omega - 4Gq - 12\gamma_2 s^2 q - 4\gamma_2 q^3=0.\tag{24}
\end{equation}
For fixed \(G>0\) and \(\Omega\to0\), we have the lowest-order results
\begin{equation}
s = \left(\frac{\Omega}{4\gamma_2}\right)^{1/3},
q = \frac{\Omega}{4G} + O(\Omega^{5/3},\Omega^3).\tag{25}
\end{equation}
Therefore, the steady-state amplitudes are
\begin{align}
\langle a\rangle_{\mathrm{ss}}
&= \left(\frac{\Omega}{4\gamma_2}\right)^{1/3}
+ \frac{\Omega}{4G}
+ \cdots, \tag{26}\\
\langle b\rangle_{\mathrm{ss}}
&= -\left(\frac{\Omega}{4\gamma_2}\right)^{1/3}
+ \frac{\Omega}{4G}
+ \cdots.\tag{27}
\end{align}

According to the error propagation formula, the measurement uncertainty of $\Omega$  obtained through homodyne detection with the measurement operator $M=b+b^\dagger$ is
\begin{equation}
\delta \Omega=\frac{\sqrt{\langle M^2\rangle-|\langle M\rangle|^2}}{|\partial_\Omega M|}\\
=6(4\Omega^2\gamma_2)^{1/3}.\tag{28}
\end{equation}

The characteristic time for reaching the steady state is determined by the eigenvalue with the smallest absolute value of its real part. (see the detail in Appendix B) :
\begin{align}
T_c=\left(\frac{\gamma_2\Omega^2}{2}\right)^{-1/3}.\tag{29}
\end{align}

The measurement uncertainty of $\Omega$ can be rewritten as
\begin{equation}
\delta \Omega=12/T_c.\tag{30}
\end{equation}
Therefore, as $\Omega$ approaches 0, the measurement uncertainty approaches zero because the time required to reach a steady state tends to infinite.
The time-domain Heisenberg limit \cite{lab27} is achieved in the limit of vanishing $\Omega$. The underlying origin is that, as $\Omega\to0$, a dark mode emerges that is immune to both collective dissipation and two-particle dissipation.
\section{Large incoherent drive}

\begin{figure}[h]
\includegraphics[scale=0.50]{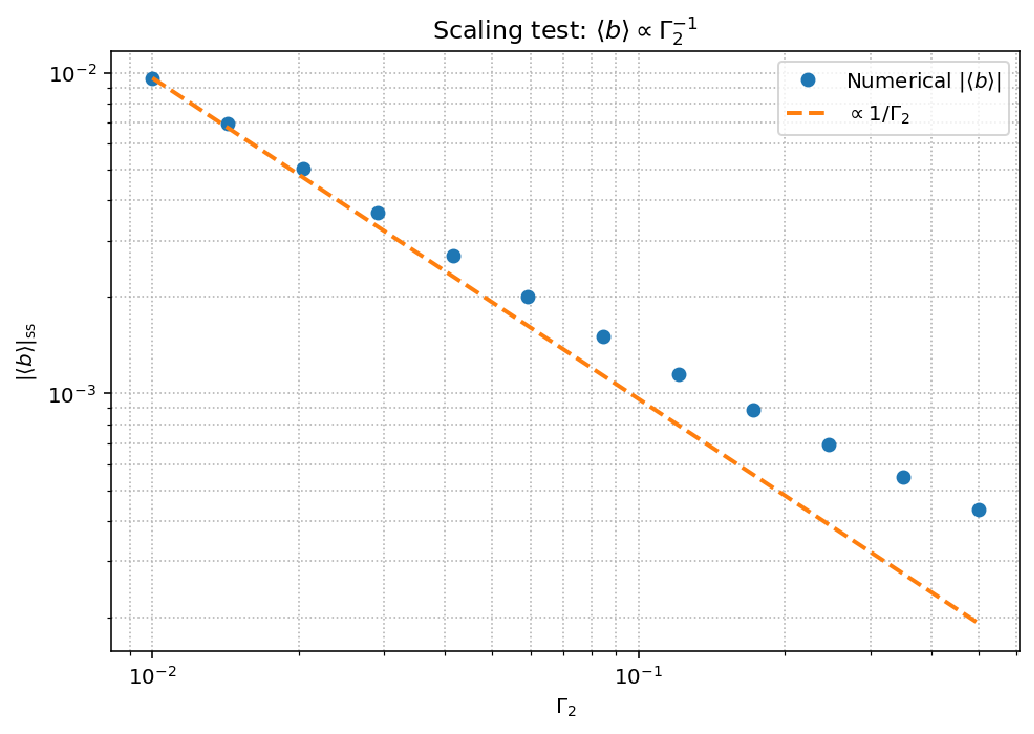}
 \caption{\label{fig.2} Comparison diagram of using Kramers-Moyal approximation and numerical calculation in the case of non-reciprocal coupling. Here, the dimensionless parameters are chosen as: $\gamma_1^+=1, \Gamma_1^+=0.70, \gamma_1^-=0.2, \Gamma_1^-=0.15, \gamma_2=0.2,\Omega=0.001, g=G=0.05$. }
\end{figure}

We assume the incoherent drive $\gamma_1^+$ and $\gamma_1^+$ are large enough, satisfying $\gamma_1^+-\gamma_1^--G\gg\gamma_2$ and $\Gamma_1^+-\Gamma_1^--G\gg\Gamma_2$. The Kramers-Moyal approximation \cite{lab28,lab29,lab30,lab31} is used in the large-occupation regime. As can be seen from Fig.~\ref{fig.2}, the result obtained by this approximation is close to the exact numerical solution when $\Gamma_2$ is small.
For weak coherent drive and weak interoscillator coupling, the analytical values can be obtained (see the detail in Appendix C)

\begin{align}
&\left\langle b^\dagger b\right\rangle_{\mathrm{ss}}
=
\frac{\Gamma_1^+-\Gamma_1^--G}{2\Gamma_2}
+
O(g^2,G^2,gG,\Omega^2),\tag{31}\\
&\langle b\rangle_{\mathrm{ss}}
=
-\frac{
(g+G)
\left(\gamma_1^+-\gamma_1^--G\right)
\left(\Gamma_1^+-\Gamma_1^--G\right)
\Omega
}{
\gamma_2\Gamma_2
\left(
3\gamma_1^+-\gamma_1^--G
\right)
\left(
3\Gamma_1^+-\Gamma_1^--G
\right)
}\times
\nonumber\\
&
\left[
1+
\dfrac{
(g^2-G^2)
\left(\gamma_1^+-\gamma_1^--G\right)
\left(\Gamma_1^+-\Gamma_1^--G\right)
}{
\gamma_2\Gamma_2
\left(
3\gamma_1^+-\gamma_1^--G
\right)
\left(
3\Gamma_1^+-\Gamma_1^--G
\right)
}
\right]^{-1}.\tag{32}
\label{eq:weak_explicit_b}
\end{align}

For the non-reciprocal coupling $G=g$, using the homodyne detection with the measurement operator $M=b+b^\dagger$, the measurement uncertainty of $\Omega$ is given by
\begin{align}
\delta\Omega|_{nr}
&\simeq
\frac{
\gamma_2\sqrt{\Gamma_2}(
3\gamma_1^+-\gamma_1^--G)(
3\Gamma_1^+-\Gamma_1^--G)
}{
4g
\left(\gamma_1^+-\gamma_1^--G\right)
\sqrt{\Gamma_1^+-\Gamma_1^--G}
}.\tag{33}
\end{align}
This result indicates that when two-particle loss rate $\Gamma_2$ approaches 0, the non-reciprocal coupling can help the measurement uncertainty approach 0.

For the reciprocal coupling $G=0$, $g>0$ and $\Gamma_2\rightarrow0$,  the measurement uncertainty of $\Omega$ is given by
\begin{align}
\delta\Omega_{r}
&\propto\frac{1}{\sqrt{\Gamma_2}}.\tag{34}
\end{align}

When the two-particle loss rate $\Gamma_2$ is very small, non-reciprocal coupling can achieve a higher improvement in measurement precision compared to reciprocal coupling. When $\Gamma_2$ is close to 0, the improvement factor $\eta$ for the measurement precision of the weak driving strength $\Omega$ is
\begin{align}
\eta=\delta\Omega_{r}/\delta\Omega|_{nr}\propto\frac{1}{\Gamma_2}.\tag{35}
\end{align}
This result indicates that in the presence of strong incoherent drive and weak two-particle loss rate, the use of non-reciprocal coupling can effectively facilitate the sensing and measurement of weakly driving signals.
\section{critical incoherent drive}
In the previous section, we have already explored the case of strong driving. In this part, we investigate the regime where incoherent drive is near its critical value.
The linear rates obey the critical condition
 \begin{equation}
 \gamma_1^+-\gamma_1^-= \Gamma_1^+-\Gamma_1^-=G. \tag{36}\label{eq:app_balance}
 \end{equation}
 The reciprocal and nonreciprocal cases are compared at fixed \(\gamma_1^+,\Gamma_1^+,\gamma_2,g,\Omega\), while \(\Gamma_2\) is varied. For \(G=g\), positivity of the specified local dissipative rates requires
 \begin{equation}
 0<g\leq \min(\gamma_1^+,\Gamma_1^+).
 \tag{37}\end{equation}
\subsection{The complete non-reciprocal coupling: G=g}
\begin{figure}[h]
\includegraphics[scale=0.50]{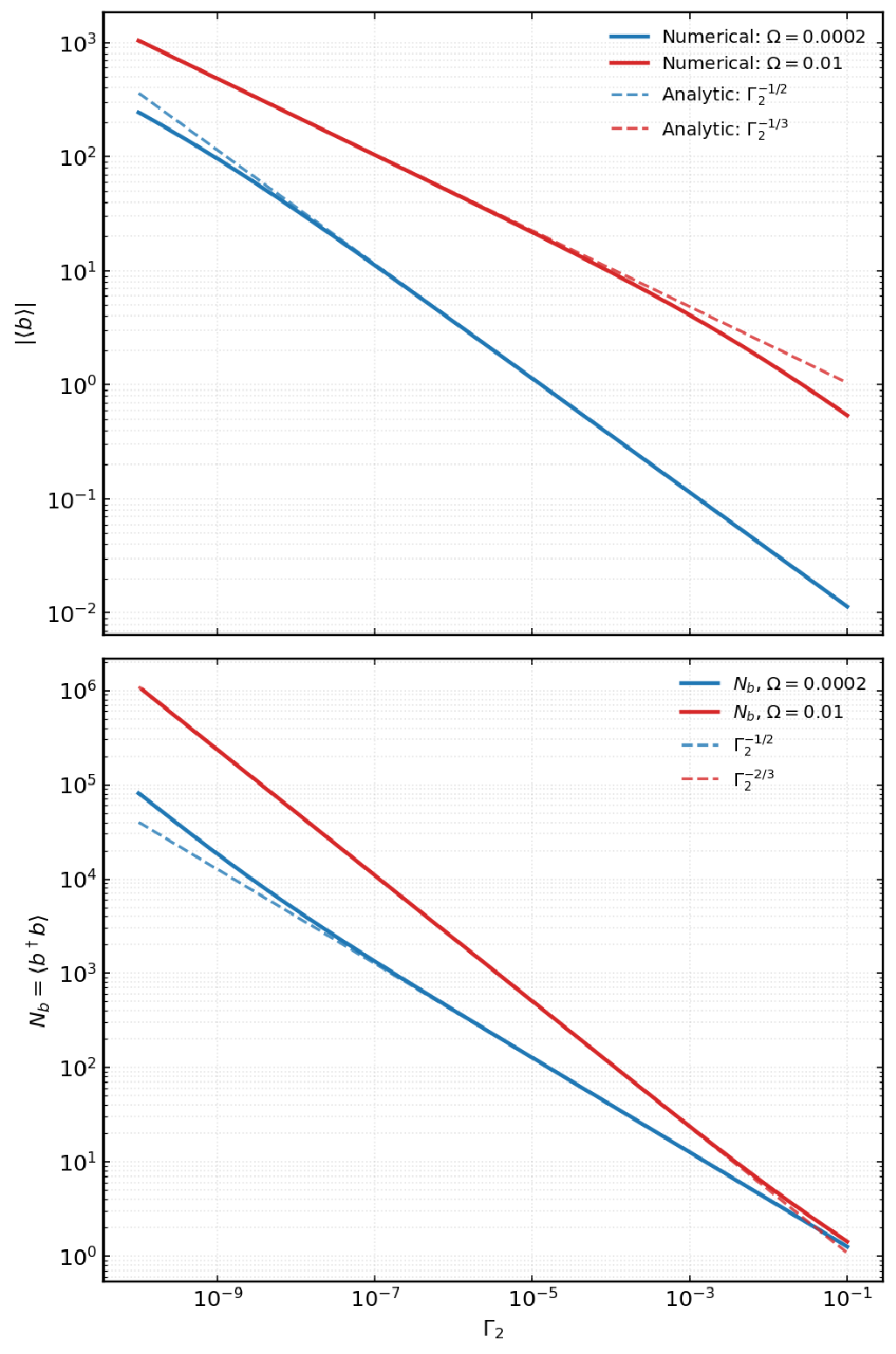}
 \caption{\label{fig.3} Evolution graph of the steady-state expectation value of model $b$ with respect to parameter $\Gamma_2$: Comparison of analytical and numerical results. Here, the dimensionless parameters are chosen as: $\gamma_1^+=1, \Gamma_1^+=0.50, \gamma_2=0.0001, g=G=0.2$. }
\end{figure}

For the complete non-reciprocal coupling, the large-occupation, time-scale-separated noise-dominated theory gives (see the detail in Appendix D)
\begin{align}
&G=g, \Omega\rightarrow0, \Gamma_2\rightarrow0:\nonumber\\
&\qquad N_b\big|_{\Omega=0}\propto\Gamma_2^{-1/2}, |\partial_\Omega \langle b\rangle_s| \propto\Gamma_2^{-1/2}. \tag{38}\label{eq:app_summary_linear}
\end{align}
At fixed nonzero drive, with nonzero upstream amplitude and finite effective noise, the asymptotic response is instead
\begin{align}
&G=g,\ \Omega\neq0, \Gamma_2\rightarrow0:\nonumber\\
&\qquad |\langle \partial_\Omega \langle b\rangle_s\rangle|\propto\Gamma_2^{-1/3}, \qquad N_b\propto\Gamma_2^{-2/3}. \tag{39}\label{eq:app_summary_fixed}
\end{align}
From Fig.~\ref{fig.3}, it can be verified that the derived relationship equations obtained from the above two scenarios are reasonable within a very narrow range of the $\Gamma_2$ value.
The distinction between the two coupling configurations is that reciprocal coupling permits downstream excitations to return to the nonlinear dissipative upstream mode, whereas the cascaded configuration removes this return pathway. The endpoint identities and the effective asymptotics should nevertheless be distinguished from a mathematical proof of stationary-state existence and of coarse-graining error bounds.
By using the homodyne detection, we can obtain the measurement uncertainty of $\Omega$ in the critical regime
\begin{align}
\delta\Omega_c
&\propto
\Gamma_2^{1/4}, \Omega\rightarrow 0,\tag{40}\\
\delta\Omega_c
&\propto
\Gamma_2^0,\ \ \ \Omega> 0.\tag{41}
\end{align}
This result indicates that only when $\Omega$ approaches 0, will the measurement uncertainty of $\Omega$ tend to 0 as $\Gamma_2$ approaches 0.
Compared with the result obtained by using a sufficiently strong incoherent drive in the previous section, the ratio of the measurement uncertainties at $\Omega=0$  and at non-zero $\Omega$ is
\begin{align}
\delta\Omega_c/\delta\Omega|_{nr}&\propto{\Gamma_2}^{1/4}, \Omega\rightarrow 0,\tag{42}\\
\delta\Omega_c/\delta\Omega|_{nr}&\propto{\Gamma_2}^{1/2}, \Omega> 0.\tag{43}
\end{align}

This indicates that with the non-reciprocal coupling, increasing the strength of incoherent driving can effectively enhance the measurement precision of weak driving intensity.

\subsection{The reciprocal coupling: G=0}
\begin{figure}[h]
\includegraphics[scale=0.50]{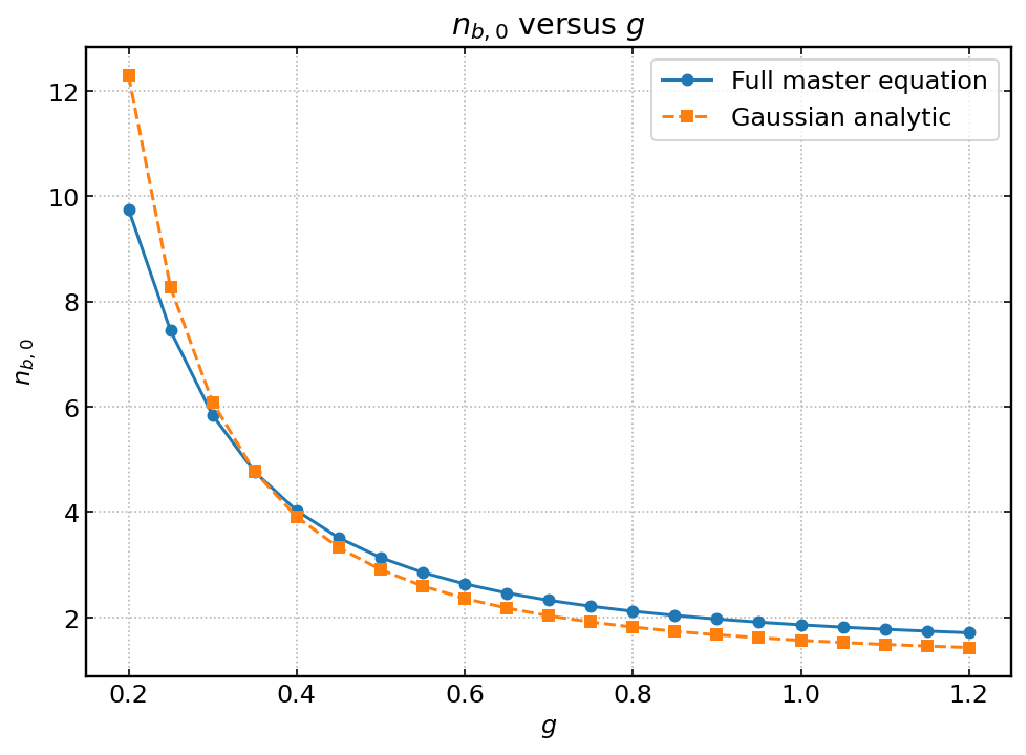}
 \caption{\label{fig.4} The graph showing the relationship between the zero-drive occupation $n_{b,0}$ and $g$: Gaussian approximation versus numerical calculation. Here, the dimensionless parameters are chosen as: $\gamma_1^+=\gamma_1^-=0.50, \Gamma_1^+=\Gamma_1^-=0.50, \gamma_2=0.2, \Omega=0, \Gamma_2=0, G=0$. }
\end{figure}

In the absence of dissipative coupling, the two systems interact only via coherent coupling.
According to the error propagation formula, the measurement uncertainty of $\Omega$ is given by (see the detail in Appendix E)
\begin{equation}
\delta\Omega =\frac{\sqrt{Var(X)}}{|\frac{\partial\langle b\rangle_{\mathrm{ss}}}{\partial\Omega}|}= \frac{|g|}{\sqrt{2}} \sqrt{n_{b,0}+\frac{1}{2}},  \tag{44}\label{eq:44}
\end{equation}
where the zero-drive ($\Omega=0$) occupation $n_{b,0}= \langle b^\dagger b\rangle_0$.
To obtain an explicit parameter-dependent expression, we adopt a zero-mean, phase-invariant Gaussian moment
closure approximation \cite{lab32}. As can be seen from Fig.~\ref{fig.4}, the approximate analytical result obtained here is essentially close to the result obtained through precise numerical calculation. Consequently,
\begin{equation}
 n_{b,0}^{\mathrm{Gauss}} = \left(1+\frac{4\Gamma^+_{1}\gamma_2}{g^2}\right) \sqrt{\frac{\gamma^+_{1}+\Gamma^+_{1}}{4\gamma_2}}. \tag{45}\label{eq:45}
\end{equation}
Substituting Eq.~\eqref{eq:45} into Eq.~\eqref{eq:44} yields
\begin{equation}
\delta\Omega_{\mathrm{Gauss}} = \frac{1}{\sqrt{2}} \left[ \left(g^2+4\Gamma^+_{1}\gamma_2\right) \sqrt{\frac{\gamma^+_{1}+\Gamma^+_{1}}{4\gamma_2}} + \frac{g^2}{2} \right]^{1/2}. \label{eq:gaussian_uncertainty}\tag{46}
\end{equation}
This result is clearly independent of $\Gamma_2$. Consequently, the measurement uncertainty does not tend to zero as $\Gamma_2$ approaches zero.
Although its measurement precision is inferior to that achieved via nonreciprocal coupling in the strong incoherent-driving regime, it still outperforms the reciprocal-coupling counterpart within the same regime.

\section{Conclusions}
In this article, we investigate a quantum sensing system composed of two vdP oscillators. In the case of weak incoherent dissipation and driving, the measurement precision of the weak driving strength is independent of the two-particle loss rate. The non-reciprocal coupling can improve the measurement precision by a factor of two compared with the reciprocal coupling. When the coherent coupling, incoherent driving and dissipation are all zero, a dark mode emerges at the zero-driving point, which enables the system to achieve a measurement precision reaching the Heisenberg limit. When incoherent drive is much larger than the two-particle loss rate, the nonreciprocal coupling enables the measurement uncertainty for the weak coherent drive to decrease as the two-particle loss rate is reduced. The non-reciprocal coupling has a greater effect on improving measurement precision than the reciprocal coupling, and this effect increases as the two-particle loss rate decreases.

The cascaded quantum vdP oscillator discussed in this article can be realized on many different experimental platforms, such as trapped ion platforms \cite{lab32,lab33,lab34}, superconducting circuits \cite{lab35,lab36,lab37}, cavity optomechanics \cite{lab38,lab39,lab40}, etc.
\section*{Acknowledgements}
This research was supported by the National Natural Science Foundation of China (Grant No. 12365001), and the Bagui Youth Top Talent Training Program.

\appendix

\section{non-Hermitian }

The non-Hermitian evolution under the no-jump approximation can be written as
\begin{align}
i\frac{d}{dt}|\psi\rangle=H_{\textmd{eff}}|\psi\rangle
\label{eq:A1}
\tag{A1}
\end{align}
where the effective Hamiltonian is
\begin{align}
&H_{\textmd{eff}}=i\Omega(a^\dagger-a)+ig(a^\dagger b-a b^\dagger)-i\{\gamma_{1}^+aa^\dagger+\Gamma_{1}^-a^\dagger a +\nonumber\\
&\gamma_{a2}a^{\dagger 2}a^2+\Gamma_{1}^+bb^\dagger+\gamma_{b1}^-b^\dagger b +\gamma_{b2}b^{\dagger 2}b^2+G(a^\dagger+b^\dagger)(a+b)\}
\label{eq:A2}
\tag{A2}
\end{align}

Using the bosonic commutation relations $aa^\dagger=a^\dagger a+1$ and $ bb^\dagger=b^\dagger b+1$, the effective Hamiltonian can be described as
\begin{align}
&H'_{\textmd{eff}}=i\Omega(a^\dagger-a)+ig(a^\dagger b-a b^\dagger)-i\{\gamma_{1}^++\gamma_{b1}^+\nonumber\\
& +\mu a^\dagger a +\nu b^\dagger b+\gamma_{a2}a^{\dagger2}a^2 +\gamma_{b2}b^{\dagger2}b^2+G(a^\dagger b+b^\dagger a)\}
\label{eq:A3}
\tag{A3}
\end{align}
where $\mu=\gamma_{1}^++\gamma_{1}^-+G $ and $\nu=\Gamma_{1}^++\Gamma_{1}^-+G$. The overall constant $\gamma_{1}^++\gamma_{b1}^+$  only changes the eigenvalues but not the eigenstates, so it can be removed.

We look for right eigenstates
\begin{equation}
H'_{\textmd{eff}}|\psi\rangle=\lambda|\psi\rangle.\label{eq:A4}
\tag{A4}
\end{equation}

In the weak-driving limit, $\lambda=O(\Omega^2)$. Therefore, when calculating the lowest-order wave-function coefficients, the terms $\lambda c_{nm}$ may be temporarily neglected.
We take the truncated wave function

\begin{align}
|\psi\rangle=&
c_{00}|00\rangle+c_{01}|01\rangle+c_{10}|10\rangle
+c_{11}|11\rangle\nonumber\\
&+c_{20}|20\rangle+c_{02}|02\rangle
+c_{12}|12\rangle+c_{21}|21\rangle
+c_{22}|22\rangle .\label{eq:A5}
\tag{A5}
\end{align}

Here, the annihilation operators satisfy
\begin{equation}
a|n,m\rangle=\sqrt n\,|n-1,m\rangle,\label{eq:A6}
\tag{A6}
\end{equation}
\begin{equation}
b|n,m\rangle=\sqrt m\,|n,m-1\rangle.\label{eq:A7}
\tag{A7}
\end{equation}
Substituting Eq.~\eqref{eq:A5} into Eq.~\eqref{eq:A4}, we obtain the coefficients
\begin{equation}
\begin{aligned}
c_{00}&=1+O(\Omega^2), c_{10}=\frac{\beta\Omega}{\Delta_1}+O(\Omega),\\
c_{01}&=-\frac{(g+G)\Omega}{\Delta_1}+O(\Omega),\\
c_{20}&,c_{11},c_{02}=O(\Omega^2),
c_{21},c_{12}=O(\Omega^3),
c_{22}=O(\Omega^4),
\end{aligned}\label{eq:A8}
\tag{A8}
\end{equation}
where $\Delta_1=\mu \nu+g^2-G^2$.
In this truncated pure-state approximation,
\begin{equation}
\langle a\rangle
=
\langle\psi|a|\psi\rangle,
\qquad
\langle b\rangle
=
\langle\psi|b|\psi\rangle.\label{eq:A9}
\tag{A9}
\end{equation}

Explicitly,
\begin{equation}
\begin{aligned}
\langle a\rangle={}&
c_{00}^*c_{10}
+\sqrt2c_{10}^*c_{20}
+c_{01}^*c_{11}
+\sqrt2c_{11}^*c_{21}\\
&+c_{02}^*c_{12}
+\sqrt2c_{12}^*c_{22},
\end{aligned}\label{eq:A10}
\tag{A10}
\end{equation}
\begin{equation}
\begin{aligned}
\langle b\rangle={}&
c_{00}^*c_{01}
+c_{10}^*c_{11}
+c_{20}^*c_{21}\\
&+\sqrt2c_{01}^*c_{02}
+\sqrt2c_{11}^*c_{12}
+\sqrt2c_{21}^*c_{22}.
\end{aligned}\label{eq:A11}
\tag{A11}
\end{equation}

At lowest order \(O(\Omega)\), we can achieve
\begin{equation}
\langle a\rangle
=
\frac{\beta\Omega}{\Delta_1}
+O(\Omega^3)\label{eq:A12}
\tag{A12}
\end{equation}
and
\begin{equation}
\langle b\rangle
=
-\frac{(g+G)\Omega}{\Delta_1}
+O(\Omega^3).\label{eq:A13}
\tag{A13}
\end{equation}

\begin{equation}
\langle b^\dagger b\rangle
=
-\frac{(g+G)^2\Omega^2}{g^4}
+O(\Omega^3).\label{eq:A14}
\tag{A14}
\end{equation}

\section{characteristic time of dark mode}

The quantum-Langevin equations of the operators $a$ and $b$ can be given by
\begin{align}
\frac{d}{dt} a
&= \Omega - G( a+b) - 2\gamma_2 a^\dagger a^2+\sqrt{2G}z_{in}(t)+\sqrt{2\gamma_2}{a^\dagger a_{in}}(t), \\
\frac{d}{dt} b
&= -G(a+b) - 2\gamma_2\langle b^\dagger b^2+\sqrt{2G}z_{in}(t)+\sqrt{2\gamma_2}{b^\dagger b_{in}}(t).
\end{align}
We express $a$ and $b$ linearly as their steady-state values plus a quantum fluctuation term: $a=\langle a\rangle+\delta a,b=\langle b\rangle+\delta b.$ The dynamical evolution of $\delta a$ and $\delta b$ can be described by
\begin{align}
\frac{d}{dt} \delta a
&= - G( \delta a+\delta b) - 2\gamma_2  \langle a\rangle^2 \delta a^\dagger-4\gamma_2 |\langle a\rangle|^2 \delta a +A_{in}(t), \\
\frac{d}{dt} \delta b
&= -G(\delta a+\delta b) - 2\gamma_2  \langle b\rangle^2 \delta b^\dagger-4\gamma_2 |\langle b\rangle|^2 \delta b+B_{in}(t),
\end{align}
where the quantum noise operators are given by
\begin{align}
A_{in}(t)=\sqrt{2G}z_{in}(t)+\sqrt{2\gamma_2}{\langle a^\dagger \rangle a_{in}}(t),\\
B_{in}(t)=\sqrt{2G}z_{in}(t)+\sqrt{2\gamma_2}{\langle b^\dagger\rangle b_{in}}(t).
\end{align}

Let us set
\[
\mathbf{r}=
\begin{pmatrix}
\delta a\\
\delta a^\dagger\\
\delta b\\
\delta b^\dagger
\end{pmatrix},
\qquad
\mathbf{F}_{\mathrm{in}}(t)=
\begin{pmatrix}
A_{\mathrm{in}}(t)\\
A_{\mathrm{in}}^\dagger(t)\\
B_{\mathrm{in}}(t)\\
B_{\mathrm{in}}^\dagger(t)
\end{pmatrix},
\]
quantum-Langevin equation is rewritten as
\[
\frac{d}{dt}\mathbf{r}
=
\mathbb{M}\mathbf{r}
+
\mathbf{F}_{\mathrm{in}}(t),\label{eq:B7}
\tag{B7}
\]

in which, the evolution matrix:\\
\begin{widetext}
\[
\mathbb{M}=\begin{pmatrix}
-4\gamma_2|\langle a\rangle|^2-G & -2\gamma_2\langle a\rangle^2 & -G & 0\\[4pt]
-2\gamma_2(\langle a\rangle^*)^2 & -4\gamma_2|\langle a\rangle|^2-G & 0 & -G\\[4pt]
-G & 0 & -4\gamma_2|\langle b\rangle|^2-G & -2\gamma_2\langle b\rangle^2\\[4pt]
0 & -G & -2\gamma_2(\langle b\rangle^*)^2 & -4\gamma_2|\langle b\rangle|^2-G
\end{pmatrix}.\label{eq:B8}
\tag{B8}
\]
\end{widetext}

As $\Omega$ approaches 0, the eigenvalue of $\mathbb{M}$ are
\begin{align}
\lambda_1&=-\left(\frac{\gamma_2\Omega^2}{2}\right)^{1/3}+O(\Omega),\tag{B9}\\
\lambda_2&=-3\left(\frac{\gamma_2\Omega^2}{2}\right)^{1/3}+O(\Omega),\tag{B10}
\end{align}
\begin{align}
\lambda_3&=-2G-\left(\frac{\gamma_2\Omega^2}{2}\right)^{1/3}+O(\Omega),\tag{B11}\\
\lambda_4&=-2G-3\left(\frac{\gamma_2\Omega^2}{2}\right)^{1/3}+O(\Omega).\tag{B12}\label{eq:B12}
\end{align}

The characteristic time for reaching the steady state is determined by the eigenvalue with the smallest absolute value:
\begin{align}
T_c=\frac{1}{|\lambda_1|}=\left(\frac{\gamma_2\Omega^2}{2}\right)^{-1/3}.\label{eq:B10}
\tag{B10}
\end{align}

\section{Weak-coupling Kramers--Moyal analysis}
\label{app:weak_km}
In this Appendix, we formulate the weak-drive response of two quantum vdP oscillators coupled via coherent exchange
and collective dissipation. We distinguish exact moment identities from the single-oscillator asymptotics and from additional closure
assumptions.

We consider two bosonic modes with annihilation operators \(a\) and \(b\).
The density matrix obeys
\begin{align}
\dot{\rho}
={}&
\left[
\Omega(a^\dagger-a)
+g(a^\dagger b-a b^\dagger),
\rho
\right]
\nonumber\\
&+
\gamma_1^+D[a^\dagger]\rho
+\gamma_1^-D[a]\rho
+\gamma_2D[a^2]\rho
\nonumber\\
&+
\Gamma_1^+D[b^\dagger]\rho
+\Gamma_1^-D[b]\rho
+\Gamma_2D[b^2]\rho
+G D[a+b]\rho ,
\label{eq:weak_master}
\end{align}
where
\begin{equation}
D[c]\rho
=
2c\rho c^\dagger
-c^\dagger c\rho
-\rho c^\dagger c .
\label{eq:weak_dissipator}
\end{equation}

We assume weak coherent drive and weak interoscillator coupling.  The
Kramers--Moyal approximation is used in the large-occupation regime, away
from the critical point where the net linear gain vanishes.

%----------------------------------------------------------------------
\subsection{Collective dissipator}
%----------------------------------------------------------------------

The collective dissipator can be decomposed as
\begin{align}
D[a+b]\rho
&=
D[a]\rho+D[b]\rho+D_{\mathrm{cr}}[a,b]\rho ,
\label{eq:weak_collective_decomposition}
\\
D_{\mathrm{cr}}[a,b]\rho
&=
2a\rho b^\dagger
+2b\rho a^\dagger
-a^\dagger b\rho
-b^\dagger a\rho
\nonumber\\
&\quad
-\rho a^\dagger b
-\rho b^\dagger a .
\label{eq:weak_cross_dissipator}
\end{align}

We define the effective local one-particle loss rates
\begin{equation}
\mu_a=\gamma_1^-+G,
\qquad
\mu_b=\Gamma_1^-+G ,
\label{eq:weak_effective_losses}
\end{equation}
and the corresponding net linear gain rates
\begin{equation}
\delta_a
=
\gamma_1^+-\mu_a
=
\gamma_1^+-\gamma_1^--G ,
\label{eq:weak_delta_a}
\end{equation}
\begin{equation}
\delta_b
=
\Gamma_1^+-\mu_b
=
\Gamma_1^+-\Gamma_1^--G .
\label{eq:weak_delta_b}
\end{equation}

The local Liouvillians are
\begin{align}
\mathcal{L}_a\rho
&=
\gamma_1^+D[a^\dagger]\rho
+\mu_aD[a]\rho
+\gamma_2D[a^2]\rho ,
\label{eq:weak_local_La}
\\
\mathcal{L}_b\rho
&=
\Gamma_1^+D[b^\dagger]\rho
+\mu_bD[b]\rho
+\Gamma_2D[b^2]\rho .
\label{eq:weak_local_Lb}
\end{align}

The term that couples the two oscillators is
\begin{equation}
\mathcal{L}_{\mathrm{c}}\rho
=
[g(a^\dagger b-a b^\dagger),\rho]
+G D_{\mathrm{cr}}[a,b]\rho .
\label{eq:weak_coupling_L}
\end{equation}

%----------------------------------------------------------------------
\subsection{Population Kramers--Moyal equation}
%----------------------------------------------------------------------

For oscillator \(a\), let
\begin{equation}
p_n^{(a)}
=
\langle n|\rho_a|n\rangle
\label{eq:weak_population_a}
\end{equation}
denote the diagonal Fock-state population.  The relevant transitions are
\begin{align}
n\longrightarrow n+1,
\qquad
W_+^{(a)}(n)
&=
2\gamma_1^+(n+1),
\label{eq:weak_gain_rate_a}
\\
n\longrightarrow n-1,
\qquad
W_-^{(a)}(n)
&=
2\mu_a n,
\label{eq:weak_loss_rate_a}
\\
n\longrightarrow n-2,
\qquad
W_2^{(a)}(n)
&=
2\gamma_2n(n-1).
\label{eq:weak_two_loss_rate_a}
\end{align}

The exact population equation is
\begin{align}
\dot p_n^{(a)}
={}&
2\gamma_1^+
\left[
n p_{n-1}^{(a)}
-(n+1)p_n^{(a)}
\right]
\nonumber\\
&+
2\mu_a
\left[
(n+1)p_{n+1}^{(a)}
-np_n^{(a)}
\right]
\nonumber\\
&+
2\gamma_2
\left[
(n+1)(n+2)p_{n+2}^{(a)}
-n(n-1)p_n^{(a)}
\right].
\label{eq:weak_population_master_a}
\end{align}

For a jump process with jump size \(\Delta n\), we define
\begin{equation}
A_k(n)
=
\sum_{\Delta n}
(\Delta n)^kW(n,\Delta n).
\label{eq:weak_KM_definition}
\end{equation}

The first two Kramers--Moyal coefficients are
\begin{align}
A_1^{(a)}(n)
&=
2\gamma_1^+(n+1)
-2\mu_a n
-4\gamma_2n(n-1),
\label{eq:weak_A1_a}
\\
A_2^{(a)}(n)
&=
2\gamma_1^+(n+1)
+2\mu_a n
+8\gamma_2n(n-1).
\label{eq:weak_A2_a}
\end{align}

The corresponding Fokker--Planck equation is
\begin{equation}
\frac{\partial p_a(n,t)}{\partial t}
=
-\frac{\partial}{\partial n}
\left[
A_1^{(a)}(n)p_a(n,t)
\right]
+
\frac{1}{2}
\frac{\partial^2}{\partial n^2}
\left[
A_2^{(a)}(n)p_a(n,t)
\right].
\label{eq:weak_FPE_a}
\end{equation}

For oscillator \(b\), the corresponding coefficients follow from
\[
\gamma_1^+\rightarrow\Gamma_1^+,
\qquad
\mu_a\rightarrow\mu_b,
\qquad
\gamma_2\rightarrow\Gamma_2.
\]

%----------------------------------------------------------------------
\subsection{Off-critical limit-cycle occupation}
%----------------------------------------------------------------------

We first assume
\begin{equation}
\delta_a>0,
\qquad
\delta_b>0,
\label{eq:weak_positive_gain}
\end{equation}
and sufficiently large occupations.

For \(n\gg1\), the drift coefficient of oscillator \(a\) becomes
\begin{equation}
A_1^{(a)}(n)
\simeq
2\delta_a n-4\gamma_2n^2.
\label{eq:weak_large_n_drift_a}
\end{equation}

The nonzero stable drift zero is given by
\begin{equation}
n_{a,*}
=
\frac{\delta_a}{2\gamma_2}.
\label{eq:weak_drift_zero_a}
\end{equation}

Linearizing around \(n_{a,*}\), with \(n=n_{a,*}+x\), yields
\begin{equation}
A_1^{(a)}(n)
\simeq
-2\delta_a x .
\label{eq:weak_linearized_drift_a}
\end{equation}

At the drift zero,
\begin{align}
A_2^{(a)}(n_{a,*})
&\simeq
2(\gamma_1^++\mu_a)n_{a,*}
+8\gamma_2n_{a,*}^2
\nonumber\\
&=
\frac{\delta_a}{\gamma_2}
\left(
3\gamma_1^+-\mu_a
\right).
\label{eq:weak_diffusion_a}
\end{align}

The resulting Gaussian approximation has variance
\begin{equation}
\operatorname{Var}(n_a)
\simeq
\frac{A_2^{(a)}(n_{a,*})}{4\delta_a}
=
\frac{3\gamma_1^+-\mu_a}{4\gamma_2}.
\label{eq:weak_variance_a}
\end{equation}

The Gaussian approximation requires
\begin{equation}
n_{a,*}^2
\gg
\operatorname{Var}(n_a),
\label{eq:weak_gaussian_condition_a}
\end{equation}
or, parametrically,
\begin{equation}
\delta_a^2
\gg
\gamma_2
\left(
3\gamma_1^+-\mu_a
\right).
\label{eq:weak_gaussian_condition_a_explicit}
\end{equation}

Therefore,
\begin{equation}
\left\langle a^\dagger a\right\rangle_0
\simeq
\frac{\delta_a}{2\gamma_2}.
\label{eq:weak_occupation_a}
\end{equation}

For oscillator \(b\), we obtain
\begin{equation}
\left\langle b^\dagger b\right\rangle_0
\simeq
\frac{\delta_b}{2\Gamma_2},
\label{eq:weak_occupation_b}
\end{equation}
and
\begin{equation}
\operatorname{Var}(n_b)
\simeq
\frac{3\Gamma_1^+-\mu_b}{4\Gamma_2}.
\label{eq:weak_variance_b}
\end{equation}

%----------------------------------------------------------------------
\subsection{Explicit local susceptibility}
%----------------------------------------------------------------------

We next calculate the local zero-frequency susceptibility in the same
large-occupation approximation.  We write
\begin{equation}
\rho_a
=
\rho_a^{(0)}
+F_a\rho_a^{(1)}
+O(F_a^2),
\qquad
\mathcal{L}_a\rho_a^{(0)}=0,
\label{eq:weak_response_expansion_a}
\end{equation}
where \(F_a\) is a weak coherent force.  The first-order correction
satisfies
\begin{equation}
\mathcal{L}_a\rho_a^{(1)}
=
-[a^\dagger-a,\rho_a^{(0)}].
\label{eq:weak_response_equation_a}
\end{equation}

Define the nearest off-diagonal matrix elements
\begin{equation}
x_n^{(a)}
=
\langle n|\rho_a^{(1)}|n-1\rangle,
\qquad
q_n^{(a)}
=
\sqrt{n}\,x_n^{(a)}.
\label{eq:weak_coherence_variables_a}
\end{equation}

The local susceptibility is
\begin{equation}
\langle a\rangle
=
\chi_a F_a,
\qquad
\chi_a
=
\sum_{n=1}^{\infty}q_n^{(a)}.
\label{eq:weak_chi_definition_a}
\end{equation}

Projecting the first-order equation onto the
\(\langle n|\cdots|n-1\rangle\) sector gives
\begin{align}
{}&
\gamma_1^+
\left[
2\sqrt{n(n-1)}\,x_{n-1}^{(a)}
-(2n+1)x_n^{(a)}
\right]
\nonumber\\
&+
\mu_a
\left[
2\sqrt{n(n+1)}\,x_{n+1}^{(a)}
-(2n-1)x_n^{(a)}
\right]
\nonumber\\
&+
\gamma_2
\left[
2(n+1)\sqrt{n(n+2)}\,x_{n+2}^{(a)}
-2(n-1)^2x_n^{(a)}
\right]
\nonumber\\
&+
\sqrt{n}
\left(
p_{n-1}^{(a)}-p_n^{(a)}
\right)=0.
\label{eq:weak_coherence_recurrence_a}
\end{align}

The source term is proportional to
\(p_{n-1}^{(a)}-p_n^{(a)}\).  This difference arises from
\[
\langle n|[a^\dagger-a,\rho_a^{(0)}]|n-1\rangle
=
\sqrt{n}\left(p_{n-1}^{(a)}-p_n^{(a)}\right).
\]

In the large-occupation limit, the recurrence relation reduces to a differential
equation for the first angular-harmonic coherence envelope \(q_a(n)\):
\begin{equation}
\mathcal{K}_a q_a(n)
-
\Lambda_a q_a(n)
=
-p_a'(n)
+\frac12p_a''(n),
\label{eq:weak_coherence_KM_a}
\end{equation}
where
\begin{equation}
\mathcal{K}_a q
=
-\partial_n\!\left[A_1^{(a)}q\right]
+
\frac12\partial_n^2\!\left[A_2^{(a)}q\right],
\label{eq:weak_coherence_operator_a}
\end{equation}
and
\begin{equation}
\Lambda_a
=
3\gamma_1^+-\mu_a.
\label{eq:weak_coherence_gap_a}
\end{equation}

The population distribution \(p_a(n)\) is the zero mode of the population
operator,
\begin{equation}
\mathcal{K}_a p_a(n)=0.
\label{eq:weak_population_zero_mode_a}
\end{equation}

A pointwise ansatz \(q_a(n)=\chi_a p_a(n)\) is not sufficient to solve Eq.~\eqref{eq:weak_coherence_KM_a}, because the driving source contains
higher Hermite components.  Instead, we retain the leading population mode and the first nonvanishing odd Hermite component generated by the
source.  Denoting the corresponding projection by \(\mathcal{P}_1\), we obtains
\begin{equation}
\mathcal{P}_1
\left[
\mathcal{K}_a q_a
-\Lambda_a q_a
\right]
=
\mathcal{P}_1
\left[
-p_a'
+\frac12p_a''
\right].
\label{eq:weak_projected_coherence_a}
\end{equation}

Using the Gaussian population distribution with
\[
\sigma_a^2
=
\frac{3\gamma_1^+-\mu_a}{4\gamma_2},
\]
the projected source and response give the closure relation
\begin{equation}
\frac{\gamma_2\Lambda_a}{\delta_a}\chi_a=1.
\label{eq:weak_chi_closure_a}
\end{equation}

Therefore, the explicit leading-order susceptibility of oscillator \(a\)
is given by
\begin{equation}
\chi_a
=
\frac{\delta_a}
{\gamma_2(3\gamma_1^+-\mu_a)}.
\label{eq:weak_explicit_chi_a}
\end{equation}

Using
\[
\delta_a=\gamma_1^+-\gamma_1^--G,
\qquad
\mu_a=\gamma_1^-+G,
\]
we obtain
\begin{equation}
\chi_a
=
\frac{
\gamma_1^+-\gamma_1^--G
}{
\gamma_2
\left(
3\gamma_1^+-\gamma_1^--G
\right)
}.
\label{eq:weak_explicit_chi_a_rates}
\end{equation}

The same calculation for oscillator \(b\) gives
\begin{equation}
\chi_b
=
\frac{\delta_b}
{\Gamma_2(3\Gamma_1^+-\mu_b)}.
\label{eq:weak_explicit_chi_b}
\end{equation}

Equivalently,
\begin{equation}
\chi_b
=
\frac{
\Gamma_1^+-\Gamma_1^--G
}{
\Gamma_2
\left(
3\Gamma_1^+-\Gamma_1^--G
\right)
}.
\label{eq:weak_explicit_chi_b_rates}
\end{equation}

The explicit formulas
\eqref{eq:weak_explicit_chi_a_rates} and
\eqref{eq:weak_explicit_chi_b_rates} require
\begin{equation}
\delta_a>0,
\qquad
\delta_b>0,
\label{eq:weak_chi_positive_gain}
\end{equation}
together with the large-occupation conditions
\begin{equation}
\frac{\delta_a}{\gamma_2}\gg1,
\qquad
\frac{\delta_b}{\Gamma_2}\gg1.
\label{eq:weak_chi_large_occupation}
\end{equation}

The above results should not be extrapolated to the critical point
\(\delta_a=\delta_b=0\), where the present (leading-order) expansion breaks down.

%----------------------------------------------------------------------
\subsection{Weak coupling between the oscillators}
%----------------------------------------------------------------------

The exact first-moment equations are
\begin{align}
\frac{d}{dt}\langle a\rangle
={}&
\Omega
+\delta_a\langle a\rangle
+(g-G)\langle b\rangle
-2\gamma_2\langle a^\dagger a^2\rangle ,
\label{eq:weak_exact_first_moment_a}
\\
\frac{d}{dt}\langle b\rangle
={}&
\delta_b\langle b\rangle
-(g+G)\langle a\rangle
-2\Gamma_2\langle b^\dagger b^2\rangle .
\label{eq:weak_exact_first_moment_b}
\end{align}

Within the local linear-response approximation, oscillator \(a\) responds
to the effective force
\begin{equation}
F_a
=
\Omega+(g-G)\langle b\rangle ,
\label{eq:weak_force_a}
\end{equation}
whereas oscillator \(b\) responds to
\begin{equation}
F_b
=
-(g+G)\langle a\rangle .
\label{eq:weak_force_b}
\end{equation}

Using the explicit local susceptibilities, we obtain
\begin{align}
\langle a\rangle
&=
\chi_a
\left[
\Omega+(g-G)\langle b\rangle
\right],
\label{eq:weak_response_equation_a}
\\
\langle b\rangle
&=
-\chi_b(g+G)\langle a\rangle .
\label{eq:weak_response_equation_b}
\end{align}

Solving the two linear equations yields
\begin{equation}
\langle a\rangle_{\mathrm{ss}}
=
\frac{\chi_a\Omega}
{1+(g^2-G^2)\chi_a\chi_b}
\label{eq:weak_resummed_a}
\end{equation}
and
\begin{equation}
\langle b\rangle_{\mathrm{ss}}
=
-\frac{(g+G)\chi_a\chi_b\Omega}
{1+(g^2-G^2)\chi_a\chi_b}
.
\label{eq:weak_resummed_b}
\end{equation}

Substituting Eqs.~\eqref{eq:weak_explicit_chi_a_rates} and
\eqref{eq:weak_explicit_chi_b_rates} into the above equations, we obtain
\begin{align}
\langle a\rangle_{\mathrm{ss}}
={}&
\frac{
\dfrac{
\left(\gamma_1^+-\gamma_1^--G\right)\Omega
}{
\gamma_2
\left(
3\gamma_1^+-\gamma_1^--G
\right)
}
}
{
1+
\dfrac{
(g^2-G^2)
\left(\gamma_1^+-\gamma_1^--G\right)
\left(\Gamma_1^+-\Gamma_1^--G\right)
}{
\gamma_2\Gamma_2
\left(
3\gamma_1^+-\gamma_1^--G
\right)
\left(
3\Gamma_1^+-\Gamma_1^--G
\right)
}
},
\label{eq:weak_explicit_a}
\\
\langle b\rangle_{\mathrm{ss}}
={}&
-\frac{
(g+G)
\left(\gamma_1^+-\gamma_1^--G\right)
\left(\Gamma_1^+-\Gamma_1^--G\right)
\Omega
}{
\gamma_2\Gamma_2
\left(
3\gamma_1^+-\gamma_1^--G
\right)
\left(
3\Gamma_1^+-\Gamma_1^--G
\right)
}
\nonumber\\
&\times
\left[
1+
\dfrac{
(g^2-G^2)
\left(\gamma_1^+-\gamma_1^--G\right)
\left(\Gamma_1^+-\Gamma_1^--G\right)
}{
\gamma_2\Gamma_2
\left(
3\gamma_1^+-\gamma_1^--G
\right)
\left(
3\Gamma_1^+-\Gamma_1^--G
\right)
}
\right]^{-1}.
\label{eq:weak_explicit_b}
\end{align}

When the coupling is completely nonreciprocal ($G=g$)
\begin{equation}
\langle a\rangle_{\mathrm{ss}}
\simeq
\frac{
\gamma_1^+-\gamma_1^--G
}{
\gamma_2
\left(
3\gamma_1^+-\gamma_1^--G
\right)
}
\Omega
\label{eq:weak_explicit_first_order_a}
\end{equation}
and
\begin{equation}
\begin{aligned}
\langle b\rangle_{\mathrm{ss}}
&\simeq
-
\frac{
(g+G)
\left(\gamma_1^+-\gamma_1^--G\right)
\left(\Gamma_1^+-\Gamma_1^--G\right)
}{
\gamma_2\Gamma_2
}
\\
&\quad\times
\frac{\Omega}{
\left(
3\gamma_1^+-\gamma_1^--G
\right)
\left(
3\Gamma_1^+-\Gamma_1^--G
\right)
}.
\end{aligned}
\label{eq:weak_explicit_first_order_b}
\end{equation}

%----------------------------------------------------------------------
\subsection{Occupation-number corrections}
%----------------------------------------------------------------------

The exact number equations are
\begin{align}
\frac{d}{dt}\langle n_a\rangle
={}&
\Omega\langle a+a^\dagger\rangle
+2\delta_a\langle n_a\rangle
+2\gamma_1^+
\nonumber\\
&-
4\gamma_2
\langle n_a(n_a-1)\rangle
+(g-G)
\langle a^\dagger b+a b^\dagger\rangle ,
\label{eq:weak_number_a}
\\
\frac{d}{dt}\langle n_b\rangle
={}&
2\delta_b\langle n_b\rangle
+2\Gamma_1^+
\nonumber\\
&-
4\Gamma_2
\langle n_b(n_b-1)\rangle
-(g+G)
\langle a^\dagger b+a b^\dagger\rangle .
\label{eq:weak_number_b}
\end{align}

The interoscillator coherence is of first order in the coupling.  It enters the
number equations multiplied by a second coupling factor.  Therefore,
\begin{equation}
\Delta n_a,\Delta n_b
=
O(g^2,G^2,gG).
\label{eq:weak_number_order}
\end{equation}

The coherent drive changes the occupations only at order \(\Omega^2\).  Hence,
to the order considered here, the steady-state occupations read
\begin{align}
\left\langle a^\dagger a\right\rangle_{\mathrm{ss}}
&=
\frac{\gamma_1^+-\gamma_1^--G}{2\gamma_2}
+
O(g^2,G^2,gG,\Omega^2),
\label{eq:weak_final_number_a}
\\
\left\langle b^\dagger b\right\rangle_{\mathrm{ss}}
&=
\frac{\Gamma_1^+-\Gamma_1^--G}{2\Gamma_2}
+
O(g^2,G^2,gG,\Omega^2).
\label{eq:weak_final_number_b}
\end{align}

%======================================================================
\section{critical regime}
This Appendix studies the response of mode \(b\) as its two-particle loss rate \(\Gamma_2\) approaches zero. We distinguish exact consequences of the Lindblad master equation from asymptotic results that hold only under large occupation, weak drive, or time-scale separation. In particular, the zero-drive linear susceptibility and the stationary response at fixed nonzero drive are governed by different limits.

We consider two bosonic modes satisfying
\begin{equation}
[a,a^\dagger]=[b,b^\dagger]=1,
\qquad
[a,b]=[a,b^\dagger]=0,
\end{equation}
with density operator evolving according to
\begin{align}
\dot{\rho}={}&[X,\rho]
+\gamma_1^+\mathcal{D}[a^\dagger]\rho
+\gamma_1^-\mathcal{D}[a]\rho
+\gamma_2\mathcal{D}[a^2]\rho
\nonumber\\
&+\Gamma_1^+\mathcal{D}[b^\dagger]\rho
+\Gamma_1^-\mathcal{D}[b]\rho
+\Gamma_2\mathcal{D}[b^2]\rho
+G\mathcal{D}[a+b]\rho ,
\label{eq:app_master}
\end{align}
where
\begin{align}
X&=\Omega(a^\dagger-a)+g(a^\dagger b-ab^\dagger),
\label{eq:app_X}\\
\mathcal{D}[L]\rho
&=2L\rho L^\dagger-\{L^\dagger L,\rho\}.
\label{eq:app_dissipator}
\end{align}
We take \(\Omega\in\mathbb{R}\) and \(g>0\). Since
\(X^\dagger=-X\), the Hamiltonian is \(H=iX\), and
\([X,\rho]=-i[H,\rho]\).

The linear rates obey the critical condition
\begin{equation}
\gamma_1^+-\gamma_1^-
=
\Gamma_1^+-\Gamma_1^-
=
G.
\label{eq:app_balance}
\end{equation}
The reciprocal and nonreciprocal cases are compared at fixed
\(\gamma_1^+,\Gamma_1^+,\gamma_2,g,\Omega\), while \(\Gamma_2\)
is varied.
For \(G=g\), positivity of the local dissipative rates requires
\begin{equation}
0<g\leq \min(\gamma_1^+,\Gamma_1^+).
\end{equation}

\subsection{Nonreciprocal coupling at \(G=g\)}
\label{app:cascade}

For \(G=g\), the Lindblad master equation can be written as
\begin{equation}
\dot{\rho}
=
(\mathcal{L}_a+\mathcal{L}_b+\mathcal{L}_{\mathrm{cas}})\rho,
\label{eq:app_cascade_master}
\end{equation}
where
\begin{align}
\mathcal{L}_a\rho
={}&[\Omega(a^\dagger-a),\rho]
+\gamma_1^+\bigl(\mathcal{D}[a]+\mathcal{D}[a^\dagger]\bigr)\rho
+\gamma_2\mathcal{D}[a^2]\rho,
\label{eq:app_La}\\
\mathcal{L}_b\rho
={}&\Gamma_1^+\bigl(\mathcal{D}[b]+\mathcal{D}[b^\dagger]\bigr)\rho
+\Gamma_2\mathcal{D}[b^2]\rho,
\label{eq:app_Lb}\\
\mathcal{L}_{\mathrm{cas}}\rho
={}&-2g\left([b^\dagger,a\rho]+[\rho a^\dagger,b]\right).
\label{eq:app_cascaded_generator}
\end{align}
The cascade term follows from
\begin{align}
&[g(a^\dagger b-ab^\dagger),\rho]
+g\bigl(\mathcal{D}[a+b]-\mathcal{D}[a]-\mathcal{D}[b]\bigr)\rho
\nonumber\\
&\hspace{2cm}
=-2g\left([b^\dagger,a\rho]+[\rho a^\dagger,b]\right).
\end{align}

Tracing over mode \(b\) gives
\begin{equation}
\dot{\rho}_a=\mathcal{L}_a\rho_a.
\end{equation}
Consequently, the stationary state of mode \(a\), denoted by
\(\rho_{a,\mathrm{ss}}\), is independent of \(\Gamma_2\). Its stationary
mean amplitude is
\begin{equation}
\alpha_a(\Omega)
\equiv
\operatorname{Tr}(a\rho_{a,\mathrm{ss}}).
\label{eq:app_alpha_a_definition}
\end{equation}
The mean force acting on mode \(b\) is
\begin{equation}
F(\Omega)
\equiv
-2g\alpha_a(\Omega).
\label{eq:app_force}
\end{equation}
For a real drive and a symmetry-preserving stationary state,
\(\alpha_a\) and \(F\) are real.

The exact mean equation for mode \(b\) is
\begin{equation}
\frac{d}{dt}\langle b\rangle
=
F-2\Gamma_2\langle b^\dagger b^2\rangle.
\label{eq:app_b_cascade}
\end{equation}
In particular, the force \(F\) is independent of \(\Gamma_2\).

\subsection{Large-occupation estimate for the upstream ($a$) mode}

An estimate of \(\alpha_a\) can be obtained from the balanced single-mode
problem
\begin{equation}
\dot{\rho}
=
\gamma_1^+
\bigl(\mathcal{D}[a]+\mathcal{D}[a^\dagger]\bigr)\rho
+\gamma_2\mathcal{D}[a^2]\rho.
\label{eq:app_single_master}
\end{equation}
The number-state transition rates are
\begin{align}
n\to n+1 &: \quad 2\gamma_1^+(n+1),\\
n\to n-1 &: \quad 2\gamma_1^+n,\\
n\to n-2 &: \quad 2\gamma_2 n(n-1).
\end{align}

For \(\gamma_2/\gamma_1^+\ll1\), the typical occupation is
\begin{equation}
n\sim\sqrt{\frac{\gamma_1^+}{\gamma_2}}\gg1.
\end{equation}
The first two Kramers--Moyal coefficients are
\begin{align}
A(n)
&=
2\gamma_1^+-4\gamma_2n^2+\mathcal{O}(n),
\\
B(n)
&=
4\gamma_1^+n+8\gamma_2n^2+\mathcal{O}(n).
\end{align}
Here the Fokker--Planck equation is written as
\begin{equation}
\partial_t p(n,t)
=
-\partial_n[A(n)p]
+\frac{1}{2}\partial_n^2[B(n)p].
\label{eq:app_fpe_convention}
\end{equation}
At large occupation, the leading terms are
\begin{equation}
A(n)\simeq2\gamma_1^+-4\gamma_2n^2,
\qquad
B(n)\simeq4\gamma_1^+n.
\label{eq:app_fpe_coefficients}
\end{equation}
The zero-current stationary solution is then given by
\begin{equation}
p_0(n)
\simeq
2\sqrt{\frac{\gamma_2}{\pi\gamma_1^+}}
\exp\left(-\frac{\gamma_2n^2}{\gamma_1^+}\right),
\qquad
n\geq0.
\label{eq:app_half_gaussian}
\end{equation}
Consequently,
\begin{align}
\langle n\rangle_0
&\simeq
\sqrt{\frac{\gamma_1^+}{\pi\gamma_2}},
\\
\langle n^2\rangle_0
&\simeq
\frac{\gamma_1^+}{2\gamma_2}.
\label{eq:app_single_moments}
\end{align}
The exact stationary number balance instead yields
\begin{equation}
\langle n(n-1)\rangle_0
=
\frac{\gamma_1^+}{2\gamma_2}.
\label{eq:app_exact_number_balance}
\end{equation}
The difference between \(\langle n^2\rangle_0\) and
\(\langle n(n-1)\rangle_0\) is subleading at large occupation.

For a weak real drive, the leading large-occupation equation of motion for the amplitude reads
\begin{equation}
d\alpha
=
\left(\Omega-2\gamma_2|\alpha|^2\alpha\right)dt
+\sqrt{\kappa_a}\,(dW_x+i\,dW_y),
\label{eq:app_single_sde}
\end{equation}
where \(dW_x,dW_y\) are independent real Wiener increments satisfying $\mathbb{E}[dW_i\,dW_j]=\delta_{ij}\,dt.$
Writing \(\alpha=x+iy\), the stationary density is
\begin{equation}
P_\Omega(x,y)
=
\frac{1}{Z_\Omega}
\exp\left[
-\frac{\gamma_2}{\gamma_1^+}(x^2+y^2)^2
+\frac{2\Omega x}{\gamma_1^+}
\right].
\label{eq:app_upstream_distribution}
\end{equation}
At zero drive, rotational symmetry gives $\langle x^2\rangle_0=\frac12\langle|\alpha|^2\rangle_0.$
Expanding to first order in \(\Omega\) yields
\begin{align}
\alpha_a
=
\langle a\rangle
&\simeq
\frac{2\Omega}{\gamma_1^+}\langle x^2\rangle_0
\nonumber\\
&\simeq
\frac{\Omega}{\sqrt{\pi\gamma_1^+\gamma_2}}.
\end{align}
Define the upstream zero-drive susceptibility by
\begin{equation}
\chi_a
\equiv
\left.
\frac{\partial\alpha_a}{\partial\Omega}
\right|_{\Omega=0}.
\label{eq:app_chi_a_definition}
\end{equation}
Then
\begin{equation}
N_a
\equiv
\langle a^\dagger a\rangle
\simeq
\sqrt{\frac{\gamma_1^+}{\pi\gamma_2}},
\qquad
\chi_a
\simeq
\frac{1}{\sqrt{\pi\gamma_1^+\gamma_2}}.
\label{eq:app_upstream_response}
\end{equation}
The weak-drive condition for this estimate is
\begin{equation}
|\Omega|
\ll
(\gamma_1^+)^{3/4}\gamma_2^{1/4}.
\label{eq:app_upstream_weak_drive}
\end{equation}

\subsection{Exact zero-drive heating rate at \(\Gamma_2=0\)}

At \(\Omega=0\) and \(\Gamma_2=0\), let
\(\rho_{a,0}\) denote the stationary upstream state, and define
\begin{equation}
M(t)
=
\operatorname{Tr}_b[b\rho(t)].
\end{equation}
Once the upstream marginal has reached \(\rho_{a,0}\),
the cascade equation gives
\begin{equation}
\dot M
=
\mathcal{L}_{a,0}M
-2g\,a\rho_{a,0}.
\end{equation}
Since
$\operatorname{Tr}(a\rho_{a,0})=0,$
the source is trace zero. The decaying stationary component is therefore
\begin{equation}
M_\perp
=
-2g\int_0^\infty
e^{\mathcal{L}_{a,0}t}
(a\rho_{a,0})\,dt.
\end{equation}
The intermode coherence is
\begin{align}
\langle a^\dagger b\rangle
&=
\operatorname{Tr}(a^\dagger M_\perp)
\nonumber\\
&=
-2g\int_0^\infty
\langle a^\dagger(t)a(0)\rangle_0\,dt,
\label{eq:app_intermode_correlation}
\end{align}
where the last equality follows from the quantum regression theorem.
Any stationary component proportional to \(\rho_{a,0}\) does not contribute
because \(\langle a^\dagger\rangle_0=0\).

At \(G=g\) and \(\Gamma_2=0\), the exact number equation for mode \(b\) is
\begin{equation}
\frac{dN_b}{dt}
=
-2g\left(
\langle a^\dagger b\rangle
+\langle ab^\dagger\rangle
\right)
+2\Gamma_1^+,
\end{equation}
where $N_b\equiv\langle b^\dagger b\rangle.$
Thus,
\begin{align}
\lim_{t\to\infty}\frac{dN_b}{dt}
&=
2\Gamma_1^+
+8g^2\operatorname{Re}
\int_0^\infty
\langle a^\dagger(t)a(0)\rangle_0\,dt
\nonumber\\
&\equiv 2K.
\end{align}
The exact zero-drive heating coefficient is therefore
\begin{equation}
K
=
\kappa_b
+
4g^2\operatorname{Re}
\int_0^\infty
\langle a^\dagger(t)a(0)\rangle_0\,dt
\label{eq:app_K_correlation}
\end{equation}
or, using \(\kappa_b=\Gamma_1^+\),
\begin{equation}
K
=
\Gamma_1^+
+
4g^2\operatorname{Re}
\int_0^\infty
\langle a^\dagger(t)a(0)\rangle_0\,dt.
\label{eq:app_K_explicit}
\end{equation}

Eq.~\eqref{eq:app_K_explicit} is an exact expression within the
quantum master equation. For the nonlinear upstream problem containing
\(\gamma_2\mathcal{D}[a^2]\), the correlation integral generally does not
reduce to an elementary expression involving only
\(\gamma_1^+,\gamma_2,g,\Gamma_1^+\). Instead, it must be evaluated from
the upstream Liouvillian.

For numerical evaluation, solve the Poisson equation
\begin{equation}
\mathcal{L}_{a,0}Y
=
-a\rho_{a,0},
\qquad
\operatorname{Tr}Y=0,
\label{eq:app_poisson}
\end{equation}
on the trace-zero subspace. Then
\begin{equation}
K
=
\Gamma_1^+
+
4g^2\operatorname{Re}\operatorname{Tr}(a^\dagger Y).
\label{eq:app_K_poisson}
\end{equation}
This gives the exact zero-drive long-time heating rate. It does not,
by itself, imply an exact white-noise description of the downstream
dynamics.

\subsection{Effective downstream stochastic description}

To obtain a local stochastic description at small positive \(\Gamma_2\),
we additionally assume that the upstream process is mixing, has a finite
correlation time \(\tau_a\), and admits a Gaussian long-time coarse
graining. We also assume a large downstream occupation. Under these
assumptions, the leading weak-drive isotropic effective equation is
\begin{equation}
d\beta
=
\left(F-2\Gamma_2|\beta|^2\beta\right)dt
+\sqrt K\,(dW_x+i\,dW_y),
\label{eq:app_downstream_sde}
\end{equation}
where
$
\beta=x+iy.
$
With this Wiener convention, additive noise alone gives
\begin{equation}
\frac{d}{dt}\mathbb{E}[|\beta|^2]=2K.
\end{equation}

Eq.~\eqref{eq:app_downstream_sde} neglects operator-ordering
corrections and noise generated by two-photon loss. The latter produces
quadrature diffusion of order \(\Gamma_2|\beta|^2\). It is subleading in
the asymptotic regimes considered below, whereas the additive diffusion
coefficient \(K\) remains finite.

At finite drive, the coarse-grained noise may be anisotropic. More
generally, with
\[
\mathbf r=(x,y)^T,
\]
one may write
\begin{equation}
d\mathbf r
=
\left[
\mathbf F-2\Gamma_2|\mathbf r|^2\mathbf r
\right]dt
+B_\Omega\,d\mathbf W,
\label{eq:app_anisotropic_sde}
\end{equation}
where
\begin{equation}
\mathbf F
=
(\operatorname{Re}F,\operatorname{Im}F)^{\mathsf T},
\qquad
Q_\Omega=B_\Omega B_\Omega^{\mathsf T}.
\end{equation}
At zero drive, phase symmetry gives $Q_0=KI_2.$
At fixed nonzero drive, neither isotropy nor equality to the zero-drive
coefficient \(K\) is assumed to hold exactly. The fixed-drive power laws
below require only that the effective diffusion matrix remain finite and
positive.

For real \(F\), the drift in Eq.~\eqref{eq:app_downstream_sde} is the
negative gradient of
\begin{equation}
V(x,y)
=
\frac{\Gamma_2}{2}(x^2+y^2)^2-Fx.
\end{equation}
The diffusion coefficient of each real quadrature in the Fokker--Planck
equation is
\begin{equation}
D_{\mathrm{quad}}=\frac K2.
\end{equation}
The stationary density is therefore
\begin{equation}
P_F(x,y)
=
\frac{1}{Z_F}
\exp\left[
-\frac{\Gamma_2}{K}(x^2+y^2)^2
+\frac{2Fx}{K}
\right].
\label{eq:app_tilted_quartic}
\end{equation}

Define the dimensionless variables
\begin{equation}
\ell
=
\left(\frac{K}{\Gamma_2}\right)^{1/4},
\qquad
h
=
\frac{2|F|}
{K^{3/4}\Gamma_2^{1/4}}.
\label{eq:app_rescaled_variables}
\end{equation}
Integration in polar coordinates gives
\begin{equation}
|\langle b\rangle|
\simeq
\ell
\frac{
\int_0^\infty s^2e^{-s^4}I_1(hs)\,ds
}{
\int_0^\infty se^{-s^4}I_0(hs)\,ds
},
\label{eq:app_beta_integral}
\end{equation}
and
\begin{equation}
N_b
\simeq
\ell^2
\frac{
\int_0^\infty s^3e^{-s^4}I_0(hs)\,ds
}{
\int_0^\infty se^{-s^4}I_0(hs)\,ds
},
\label{eq:app_nb_integral}
\end{equation}
where \(I_\nu\) is a modified Bessel function. For real nonzero \(F\),
\(\langle b\rangle\) has the same sign as \(F\).

At \(F=0\), the effective distribution of
\(n=|\beta|^2\) is
\begin{equation}
p_b(n)
\simeq
2\sqrt{\frac{\Gamma_2}{\pi K}}
\exp\left(-\frac{\Gamma_2n^2}{K}\right),
\qquad
n\geq0.
\label{eq:app_downstream_number_distribution}
\end{equation}
Consequently,
\begin{equation}
N_b
\simeq
\sqrt{\frac{K}{\pi\Gamma_2}}.
\label{eq:app_noise_number}
\end{equation}

Expanding Eq.~\eqref{eq:app_tilted_quartic} to first order in \(F\) gives
\begin{align}
\langle b\rangle
&\simeq
\frac{2F}{K}\langle x^2\rangle_{F=0}
=
\frac{F}{K}\langle|\beta|^2\rangle_{F=0}
\nonumber\\
&=
\frac{F}{\sqrt{\pi K\Gamma_2}}
=
-\frac{2g\alpha_a}{\sqrt{\pi K\Gamma_2}}.
\label{eq:app_linear_beta}
\end{align}
The downstream zero-drive susceptibility is defined by
\begin{equation}
\chi_b
\equiv
\left.
\frac{\partial\langle b\rangle}{\partial\Omega}
\right|_{\Omega=0}.
\label{eq:app_chi_b_definition}
\end{equation}
Using \(F=-2g\alpha_a\), one obtains
\begin{equation}
\chi_b
\simeq
-\frac{2g\chi_a}
{\sqrt{\pi K\Gamma_2}}.
\label{eq:app_chi_b}
\end{equation}
The linear-response result requires
\begin{equation}
|F|
\ll
K^{3/4}\Gamma_2^{1/4}.
\label{eq:app_downstream_weak_force}
\end{equation}

\subsection{Fixed nonzero drive}

Now keep \(\Omega\neq0\) fixed while taking
\(\Gamma_2\to0^+\). Since \(\alpha_a\) is independent of \(\Gamma_2\),
the force $F=-2g\alpha_a$
is fixed and nonzero, provided \(\alpha_a\neq0\). The weak-force condition
in Eq.~\eqref{eq:app_downstream_weak_force} therefore eventually fails.

In this regime, the stationary distribution concentrates near the
deterministic fixed point
\begin{equation}
F
=
2\Gamma_2|\beta_0|^2\beta_0.
\label{eq:app_fixed_point}
\end{equation}
Thus,
\begin{equation}
\beta_0
=
\frac{F}{|F|}
\left(\frac{|F|}{2\Gamma_2}\right)^{1/3}.
\end{equation}
For \(\Omega>0\), \(\alpha_a>0\), and hence \(F=-2g\alpha_a<0\),
\begin{equation}
\langle b\rangle
\simeq
-\left(\frac{g\alpha_a}{\Gamma_2}\right)^{1/3},
\qquad
N_b
\simeq
\left(\frac{g\alpha_a}{\Gamma_2}\right)^{2/3}.
\label{eq:app_fixed_drive_scaling}
\end{equation}
More generally,
\begin{equation}
|\langle b\rangle|
\propto
\Gamma_2^{-1/3},
\qquad
N_b\propto
\Gamma_2^{-2/3}.
\label{eq:app_fixed_drive_powers}
\end{equation}

\subsection{Homodyne measurement of the drive amplitude}
\label{app:homodyne}

We now consider the estimation of the real drive amplitude \(\Omega\)
from a homodyne measurement of mode \(b\). The discussion below uses the
same large-occupation and effective stochastic description as above.
Quantum corrections of order unity, including the vacuum contribution
to the quadrature variance, are subleading in the critical
large-occupation regime.

\subsubsection{Homodyne quadrature and error propagation}

For a local-oscillator phase \(\theta\), define
\begin{equation}
X_\theta
=
\frac{be^{-i\theta}+b^\dagger e^{i\theta}}{\sqrt2}.
\label{eq:app_homodyne_quadrature}
\end{equation}
For a stationary state depending on \(\Omega\), an unbiased estimator
based on a single quadrature sample obeys, to leading order,
\begin{equation}
\Delta\Omega_\theta^{(1)}
=
\frac{\sqrt{\operatorname{Var}(X_\theta)}}
{\left|\partial_\Omega\langle X_\theta\rangle\right|}.
\label{eq:app_single_shot_error}
\end{equation}

The optimal homodyne angle is the one for which the measured quadrature
is parallel to the response vector
\[
\partial_\Omega\langle b\rangle
=
|\partial_\Omega\langle b\rangle|e^{i\theta_{\mathrm{opt}}}.
\]
Thus,
\begin{equation}
X_{\mathrm{opt}}
=
\frac{
be^{-i\theta_{\mathrm{opt}}}
+b^\dagger e^{i\theta_{\mathrm{opt}}}
}{\sqrt2}.
\label{eq:app_optimal_quadrature}
\end{equation}

For a real drive and a symmetry-preserving stationary state, both
\(\langle b\rangle\) and
\(\partial_\Omega\langle b\rangle\) are real. Therefore one may choose
\begin{equation}
X=X_0
=
\frac{b+b^\dagger}{\sqrt2}.
\label{eq:app_real_quadrature}
\end{equation}
In the effective classical description \(\beta=x+iy\), we can obtain $X\simeq\sqrt2\,x.$

\subsubsection{Zero-drive linear-response sensitivity}

At zero drive, the downstream stationary distribution is isotropic.
Hence,
\begin{equation}
\langle X\rangle_{\Omega=0}=0,
\qquad
\operatorname{Var}(X)_{\Omega=0}
\simeq
2\langle x^2\rangle_0
=
N_b.
\label{eq:app_zero_drive_quadrature_variance}
\end{equation}
Using $N_b\simeq\sqrt{\frac{K}{\pi\Gamma_2}},$
and
$
\left.
\frac{\partial\langle X\rangle}{\partial\Omega}
\right|_{\Omega=0}
=
\sqrt2\,\chi_b,
$
we obtain
\begin{equation}
\Delta\Omega_{\mathrm{hom}}^{(1)}
\simeq
\frac{\sqrt{N_b}}
{\sqrt2|\chi_b|}.
\label{eq:app_homodyne_error_zero_drive_general}
\end{equation}
Using Eq.~\eqref{eq:app_chi_b},
\begin{equation}
\Delta\Omega_{\mathrm{hom}}^{(1)}
\simeq
\frac{\pi^{1/4}K^{3/4}\Gamma_2^{1/4}}
{2\sqrt2\,g|\chi_a|}.
\label{eq:app_homodyne_error_zero_drive}
\end{equation}
Therefore,
\begin{equation}
\Delta\Omega_{\mathrm{hom}}^{(1)}
\propto
\Gamma_2^{1/4}.
\label{eq:app_single_shot_scaling}
\end{equation}

This result is valid only inside the zero-drive linear-response window
\begin{equation}
|F|
\ll
K^{3/4}\Gamma_2^{1/4}.
\label{eq:app_homodyne_linear_condition}
\end{equation}
For fixed nonzero \(\Omega\), this condition eventually fails as
\(\Gamma_2\to0^+\).

\subsubsection{Fixed nonzero drive}

We next consider fixed \(\Omega\neq0\), for which \(F\neq0\) is independent
of \(\Gamma_2\). Let
\begin{equation}
r_0
=
|\beta_0|
=
\left(\frac{|F|}{2\Gamma_2}\right)^{1/3}.
\end{equation}
The mean amplitude is aligned with the force. The optimal homodyne
quadrature is therefore the quadrature parallel to \(F\).

The longitudinal fluctuation relaxation rate is
\begin{equation}
\lambda_\parallel
=
6\Gamma_2r_0^2.
\end{equation}
Since the diffusion coefficient of the longitudinal real quadrature is
\(K/2\), its variance is
\begin{equation}
\operatorname{Var}(\delta x)
\simeq
\frac{K}{12\Gamma_2r_0^2}.
\end{equation}
Therefore,
\begin{equation}
\operatorname{Var}(X_\parallel)
\simeq
2\operatorname{Var}(\delta x)
=
\frac{K}{6\Gamma_2r_0^2}.
\label{eq:app_fixed_drive_quadrature_variance}
\end{equation}

The mean quadrature is
\begin{equation}
\langle X_\parallel\rangle
\simeq
\sqrt2\,r_0.
\end{equation}
Since \(F\) depends on \(\Omega\),
\begin{equation}
\frac{\partial r_0}{\partial\Omega}
=
\frac{r_0}{3F}
\frac{\partial F}{\partial\Omega},
\label{eq:app_r0_derivative}
\end{equation}
with the appropriate real sign convention. In the weak-drive upstream
approximation,
\begin{equation}
\frac{\partial F}{\partial\Omega}
=
-2g\chi_a.
\label{eq:app_force_susceptibility}
\end{equation}

The single-sample homodyne uncertainty is therefore
\begin{align}
\Delta\Omega_\parallel^{(1)}
&\simeq
\frac{
\sqrt{K/(6\Gamma_2r_0^2)}
}{
\sqrt2\,|\partial_\Omega r_0|
}
\nonumber\\
&=
\frac{\sqrt K}
{\sqrt{12\Gamma_2}\,r_0|\partial_\Omega r_0|}.
\label{eq:app_fixed_drive_single_error}
\end{align}
Using Eq.~\eqref{eq:app_r0_derivative},
\begin{equation}
\Delta\Omega_\parallel^{(1)}
\simeq
\frac{|F|\sqrt K}
{2\sqrt{3\Gamma_2}\,
r_0^2|\partial_\Omega F|}.
\label{eq:app_fixed_drive_single_error_force}
\end{equation}
Since \(r_0^2\propto\Gamma_2^{-2/3}\),
\begin{equation}
\Delta\Omega_\parallel^{(1)}
\propto
\Gamma_2^{1/6}.
\label{eq:app_fixed_drive_single_scaling}
\end{equation}

The fixed-drive relaxation time scales as
\begin{equation}
\tau_\parallel
\sim
(\Gamma_2r_0^2)^{-1}
\sim
\Gamma_2^{-1/3}|F|^{-2/3}.
\label{eq:app_fixed_drive_correlation_time}
\end{equation}
Consequently, in the long-measurement regime $T\gg\tau_\parallel(\Gamma_2),$
one obtains
\begin{equation}
\Delta\Omega_\parallel(T)
\simeq
\sqrt{\frac{2\tau_\parallel}{T}}\,
\Delta\Omega_\parallel^{(1)}
\propto
\Gamma_2^0.
\label{eq:app_fixed_drive_finite_time_scaling}
\end{equation}
Again, the \(\Gamma_2^0\) result is a long-time scaling statement and
cannot be extrapolated to arbitrarily small \(\Gamma_2\) at strictly fixed
finite \(T\).
\section{Homodyne estimation of $\Omega$ in the limit $G=0$ and $\Gamma_2\to0$}

We consider the estimation of the real driving amplitude $\Omega$ using homodyne measurements of mode $b$. An exact displacement relation can be established at $G=0$ and $\Gamma_2=0$. This relation determines the mean response exactly, whereas an explicit expression for the steady-state noise in terms of the model parameters requires an additional approximation.

\subsection{Model and assumptions}

The Lindblad master equation at $G=0$ is
\begin{align}
\dot{\rho} ={}&-i[H,\rho] +\gamma^+_{1}\mathcal{D}[a]\rho +\gamma^+_{1}\mathcal{D}[a^\dagger]\rho +\gamma_2\mathcal{D}[a^2]\rho \nonumber\\
&+\kappa_b\mathcal{D}[b]\rho +\kappa_b\mathcal{D}[b^\dagger]\rho +\Gamma_2\mathcal{D}[b^2]\rho, \label{eq:master}
\end{align}
with
\begin{equation}
H=i\Omega(a^\dagger-a) +ig(a^\dagger b-ab^\dagger), \label{eq:H}
\end{equation}
where $g\neq0$ is real and
\begin{equation}
\mathcal{D}[L]\rho = 2L\rho L^\dagger -L^\dagger L\rho -\rho L^\dagger L. \label{eq:dissipator}
\end{equation}
Here $\gamma_2>0$ denotes the two-particle loss rate of mode $a$.

We assume that the zero-drive steady state at $\Gamma_2=0$ exists, has finite second moments, and is invariant under a common phase rotation of the two modes. A unique steady state necessarily inherits this symmetry.

\subsection{Exact displacement relation at $\Gamma_2=0$}

Introduce the displaced mode
\begin{equation}
c=b+\frac{\Omega}{g}, \qquad b=c-\frac{\Omega}{g}. \label{eq:shift}
\end{equation}
Substituting Eq.~\eqref{eq:shift} into Eq.~\eqref{eq:H} gives
\begin{align}
H &= i\Omega(a^\dagger-a)+ig\left[a^\dagger\left(c-\frac{\Omega}{g}\right)-a\left(c^\dagger-\frac{\Omega}{g}\right)\right] \nonumber\\
&=ig(a^\dagger c-ac^\dagger). \label{eq:shiftedH}
\end{align}
Thus, the drive disappears from the Hamiltonian in the displaced variables.

The balanced linear gain and loss of mode $b$ are also invariant under this displacement. For real $s=\Omega/g$, direct expansion gives
\begin{align}
\mathcal{D}[c-s]\rho &=\mathcal{D}[c]\rho+s[c^\dagger-c,\rho],\\
\mathcal{D}[c^\dagger-s]\rho &=\mathcal{D}[c^\dagger]\rho+s[c-c^\dagger,\rho].
\end{align}
Consequently,
\begin{equation}
\mathcal{D}[b]\rho+\mathcal{D}[b^\dagger]\rho = \mathcal{D}[c]\rho+\mathcal{D}[c^\dagger]\rho. \label{eq:dissipative_invariance}
\end{equation}
At $\Gamma_2=0$, the entire master equation expressed in terms of $a$ and $c$ is therefore identical to the zero-drive master equation. Let $\rho_0$ denote its zero-drive steady state. The corresponding driven steady state is
\begin{equation}
 \rho_{\mathrm{ss}}(\Omega) = D_b\left(-\frac{\Omega}{g}\right) \rho_0 D_b^\dagger\left(-\frac{\Omega}{g}\right),  \label{eq:displaced_ss}
\end{equation}
where
\begin{equation}
D_b(\alpha)=\exp(\alpha b^\dagger-\alpha^*b).
\end{equation}
Eq.~\eqref{eq:displaced_ss} is exact and does not require a Gaussian approximation. It relies on the balanced linear gain and loss of mode $b$ and on $\Gamma_2=0$.

At zero drive, the master equation is invariant under
\begin{equation}
a\mapsto ae^{i\phi}, \qquad b\mapsto be^{i\phi}.
\end{equation}
For a phase-invariant steady state,
\begin{equation}
\langle b\rangle_0=0, \qquad \langle b^2\rangle_0=0.
\end{equation}
Define the zero-drive occupation
\begin{equation}
n_{b,0}=\langle b^\dagger b\rangle_0.
\end{equation}
Using Eq.~\eqref{eq:displaced_ss}, we obtain
\begin{equation}
\langle b\rangle_{\mathrm{ss}}=-\frac{\Omega}{g}, \qquad
\frac{\partial\langle b\rangle_{\mathrm{ss}}}{\partial\Omega}=-\frac{1}{g}. \label{eq:exact_response}
\end{equation}
The second moments are
\begin{equation}
\langle b^\dagger b\rangle_{\mathrm{ss}} = n_{b,0}+\frac{\Omega^2}{g^2}, \qquad
\langle b^2\rangle_{\mathrm{ss}} = \frac{\Omega^2}{g^2}.
\end{equation}
The displacement contribution to the occupation must therefore be distinguished from the fluctuation noise.

Consider the homodyne quadrature
\begin{equation}
X_\theta = \frac{be^{-i\theta}+b^\dagger e^{i\theta}}{\sqrt{2}}.
\end{equation}
Its mean and response are
\begin{equation}
\langle X_\theta\rangle_{\mathrm{ss}} = -\frac{\sqrt{2}\Omega}{g}\cos\theta, \qquad
\frac{\partial\langle X_\theta\rangle_{\mathrm{ss}}}{\partial\Omega} = -\frac{\sqrt{2}}{g}\cos\theta.
\end{equation}
A displacement does not alter the centered fluctuations. Since the zero-drive steady state is phase invariant, it follows that
\begin{equation}
\operatorname{Var}(X_\theta) = n_{b,0}+\frac{1}{2}.  \label{eq:quadrature_variance}
\end{equation}
The variance is independent of both $\Omega$ and $\theta$.

The optimal phase for estimation based on the mean quadrature is therefore
\begin{equation}
\theta_{\mathrm{opt}}=0 \pmod{\pi}.
\end{equation}
Accordingly, we measure
\begin{equation}
X = \frac{b+b^\dagger}{\sqrt{2}}.
\end{equation}

According to the error propagation formula, the measurement uncertainty of $\Omega$ is given by
\begin{equation}
\delta\Omega =\frac{\sqrt{Var(X)}}{|\frac{\partial\langle b\rangle_{\mathrm{ss}}}{\partial\Omega}|}= \frac{|g|}{\sqrt{2}} \sqrt{n_{b,0}+\frac{1}{2}}.  \label{eq:ideal_uncertainty}
\end{equation}
This result does not require Gaussian-distributed measurement outcomes.

To express $n_{b,0}$ in terms of the system parameters, consider the zero-drive moment equations at $\Gamma_2=0$. Define
\begin{equation}
n_a=\langle a^\dagger a\rangle_0, \qquad n_b=\langle b^\dagger b\rangle_0=n_{b,0}, \qquad C=\langle a^\dagger b\rangle_0.
\end{equation}
The master equation gives
\begin{align}
\dot n_a &= 2\gamma^+_{1} -4\gamma_2\langle a^{\dagger 2}a^2\rangle +2g\,\operatorname{Re}C, \label{eq:na}\\
\dot n_b &= 2\Gamma^+_{1}-2g\,\operatorname{Re}C, \label{eq:nb}\\
\dot C &= g(n_b-n_a) -2\gamma_2\langle a^{\dagger 2}ab\rangle. \label{eq:C}
\end{align}
At steady state, Eqs.~\eqref{eq:na} and \eqref{eq:nb} imply
\begin{equation}
\operatorname{Re}C=\frac{\Gamma^+_{1}}{g}, \qquad \langle a^{\dagger 2}a^2\rangle_0 = \frac{\gamma^+_{1}+\Gamma^+_{1}}{2\gamma_2}.  \label{eq:exact_moments}
\end{equation}
However, Eq.~\eqref{eq:C} contains a fourth-order correlation. The second-order moment equations are therefore not closed, and Eq.~\eqref{eq:exact_moments} alone does not determine $n_{b,0}$.

\subsection{Gaussian moment-closure approximation}

To obtain an explicit parameter-dependent expression, we adopt a zero-mean, phase-invariant Gaussian moment closure:
\begin{equation}
\langle a^{\dagger 2}a^2\rangle \simeq 2n_a^2, \qquad
\langle a^{\dagger 2}ab\rangle \simeq 2n_a C.
\end{equation}
The first relation and Eq.~\eqref{eq:exact_moments} give
\begin{equation}
n_a^{\mathrm{Gauss}} = \sqrt{\frac{\gamma^+_{1}+\Gamma^+_{1}}{4\gamma_2}}.
\end{equation}
The steady-state equation for $C$ becomes
\begin{equation}
g(n_b-n_a)-4\gamma_2n_aC=0.
\end{equation}
Within this closure, $C$ is real and
\begin{equation}
C^{\mathrm{Gauss}}=\frac{\Gamma^+_{1}}{g}.
\end{equation}
Consequently,
\begin{equation}
 n_{b,0}^{\mathrm{Gauss}} = \left(1+\frac{4\Gamma^+_{1}\gamma_2}{g^2}\right) \sqrt{\frac{\gamma^+_{1}+\Gamma^+_{1}}{4\gamma_2}}. \label{eq:gaussian_nb}
\end{equation}
Substituting Eq.~\eqref{eq:gaussian_nb} into Eq.~\eqref{eq:ideal_uncertainty} yields
\begin{equation}
\delta\Omega_{\mathrm{Gauss}} = \frac{1}{\sqrt{2}} \left[ \left(g^2+4\Gamma^+_{1}\gamma_2\right) \sqrt{\frac{\gamma^+_{1}+\Gamma^+_{1}}{4\gamma_2}} + \frac{g^2}{2} \right]^{1/2}. \label{eq:gaussian_uncertainty}
\end{equation}
For equal linear rates, $\gamma^+_{1}=\Gamma^+_{1}=\kappa$, this reduces to
\begin{equation}
\delta\Omega_{\mathrm{Gauss}} = \frac{1}{\sqrt{2}} \left[ \left(g^2+4\kappa\gamma_2\right) \sqrt{\frac{\kappa}{2\gamma_2}} +\frac{g^2}{2\eta} \right]^{1/2}. \label{eq:symmetric_uncertainty}
\end{equation}
The limit $g\to0$ can be singular and must not be inferred by directly extending the displacement argument to $g=0$.


\begin{thebibliography}{9}
\vspace{6mm}
\bibitem{lab1}A. de O. Junior, A. L. Andersen, B. L. Larsen, S. W. Moore, D. Markham, M. Takeoka, J. B. Brask, and U. L. Andersen, PRX Quantum 7, 033036 (2026).
\bibitem{lab2}J. H. Iacoponi, M. Rademacher, and T. S. Monteiro, Nanoparticle arrays levitated in a cavity for quantum sensing, Phys. Rev. Research 8, 033286 (2026).
\bibitem{lab3}J. Bate, A. Hamann, M. Canteri, A. Winkler, Z. X. Koong, V. Krutyanskiy, W. D\"{u}r, and B. P. Lanyon, Experimental Distributed Quantum Sensing in a Noisy Environment, Phys. Rev. Lett. 135, 220801 (2025).
\bibitem{lab4}L. Xiao, S. Sarkar, K. Wang, A. Bayat, and P. Xue, Observation of Criticality-Enhanced Quantum Sensing in Nonunitary Quantum Walks, Phys. Rev. Lett. 136, 060802 (2026).
\bibitem{lab5}L. Labarca, S. Turcotte, A. Blais, and B. Royer, Quantum Sensing of Displacements with Stabilized Gottesman-Kitaev-Preskill States, PRX Quantum 7, 020301 (2026).
\bibitem{lab6}Y. Feng, Z. Zeng, J. Cheng, Z. You, H. Lu, Z. Yan, X. Jia, C. Xie, and K. Peng, Quantum-Enhanced Interferometer for Multiphase Sensing, Phys. Rev. Lett. 135, 183602 (2025).
 \bibitem{lab7}Q. Cao, G. T. Genov, Y. Chu, J. Cai, Y. Liu, A. Retzker, and F. Jelezko, Overcoming Frequency Resolution Limits Using a Solid-State Spin Quantum Sensor, Phys. Rev. Lett. 135, 250806 (2025).
\bibitem{lab8}C. L. Degen, F. Reinhard, and P. Cappellaro, Quantum sensing, Rev. Mod. Phys. 89, 035002 (2017).
\bibitem{lab9}A. A. Clerk, M. H. Devoret, S. M. Girvin, F. Marquardt, and R. J. Schoelkopf, Introduction to quantum noise, measurement, and amplification, Rev. Mod. Phys. 82, 1155 (2010).
\bibitem{lab10}M. Aspelmeyer, T. J. Kippenberg, and F. Marquardt, Cavity optomechanics, Rev. Mod. Phys. 86, 1391 (2014).
\bibitem{lab11}H.-P. Breuer, and F. Petruccione, The Theory of Open Quantum Systems (Oxford University Press, Oxford, 2002).
\bibitem{lab12}B. van der Pol, On relaxation-oscillations, The London, Edinburgh, and Dublin Philosophical Magazine and Journal of Science, 2(11), 978 (1926).
\bibitem{lab13}V. M. Egu\'{\i}luz, M. Ospeck, Y. Choe, A. J. Hudspeth, M. O. Magnasco. Essential nonlinearities in hearing. Phys. Rev. Lett., 84, 5232 (2000).
\bibitem{lab14}A. J. Hudspeth, Making an effort to listen: Mechanical amplification in the ear, Neuron, 59, 530 (2008).
\bibitem{lab15}N. L\"{o}rch, E. Amitai, A. Nunnenkamp, and C. Bruder, Genuine quantum signatures in synchronization of anharmonic self-oscillators, Phys. Rev. Lett., 117, 073601 (2016).
\bibitem{lab16}M. R. Hush, W. Li, S. Genway, I. Lesanovsky, A. D. Armour. Spin correlations as a probe of quantum synchronization in trapped-ion phonon lasers, Phys. Rev. A, 91, 061401(R) (2015).
\bibitem{lab17}D. Xie, and C. Xu, Quantum sensing with nonreciprocal couplings, Phys. Rev. Applied, 22, 064072 (2024).
\bibitem{lab18}A. Metelmann, A. A. Clerk. Nonreciprocal photon transmission and amplification via reservoir engineering, Phys. Rev. X, 5, 021025 (2015).
\bibitem{lab19}N. Thomas and M. Senthilvelan, Normal-mode-splitting-induced synchronization blockade in coupled quantum van der Pol oscillators, Phys. Rev. A 113, L020202 (2026).
\bibitem{lab20}N. Thomas and M. Senthilvelan, Second-harmonic phase locking and synchronization blockade in quadratically coupled driven quantum Van der Pol oscillators, Phys. Rev. A 114, 032215 (2026)
\bibitem{lab21}M. Kumar and B. K. Agarwalla, Understanding synchronization between quantum self-sustained oscillators through coherence generation, Phys. Rev. A 113, 012218 (2026). \bibitem{lab22}T.Kehrer, and C. Bruder, Quantum synchronization blockade induced by nonreciprocal coupling,  Phys. Rev. A 112, 012223 (2025).
  \bibitem{lab23}T. E. Lee and H. R. Sadeghpour, Quantum synchronization of quantum van der Pol oscillators with trapped ions,
Phys. Rev. Lett. 111, 234101 (2013).
\bibitem{lab24}S. Dutta, and N. R. Cooper, Critical Response of a Quantum van der Pol Oscillator,  Phys. Rev. Lett. 123, 250401 (2019).
\bibitem{lab25}J. Dalibard, Y. Castin, and K. M{\o}lmer, Wave-function approach to dissipative processes in quantum optics, Phys. Rev. Lett. 68, 580 (1992).
\bibitem{lab26}H. J. Carmichael, An Open Systems Approach to Quantum Optics, (Springer, Berlin, 1993).
\bibitem{lab27}V. Giovannetti, S. Lloyd, L. Maccone, Quantum-enhanced measurements: beating the standard quantum limit, Science 306, 1330 (2004).
\bibitem{lab28}J. E. Moyal, Stochastic processes and statistical physics, J. R. Stat. Soc. B 11, 150 (1949).
\bibitem{lab29}D. F. Walls and G. J. Milburn, \textit{Quantum Optics}, 2nd ed. (Springer, Berlin, 2008).
\bibitem{lab30}C. W. Gardiner, \textit{Handbook of Stochastic Methods}, 4th ed. (Springer, Berlin, 2009).
\bibitem{lab31}C. W. Gardiner, P. Zoller, Quantum Noise: A Handbook of Markovian and Non-Markovian Quantum Stochastic Methods with Applications to Quantum Optics, 3rd ed., Springer (2004).
\bibitem{lab32}C. Weedbrook, S. Pirandola, R. Garc\'{\i}a-Patr\'{o}n, N. J. Cerf, T. C. Ralph, J. H. Shapiro, and S. Lloyd, Gaussian quantum information, Rev. Mod. Phys. 84, 621 (2012).
\bibitem{lab33}T. E. Lee and H. R. Sadeghpour, Quantum Synchronization of Quantum van der Pol Oscillators with Trapped Ions,
Phys. Rev. Lett. 111, 234101 (2013).
\bibitem{lab34}Y. Li, Z. Xie, X. Yang, Y. Li, X. Zhao, X. Cheng, X. Peng, J. Li, E. Lutz, Y. Lin, and J. Du, Experimental realization and synchronization of a quantum van der Pol oscillator,
Sci. Adv. 11, eady5649 (2025).
\bibitem{lab35}J. Liu, Q. Wu, J. E. Moore, H. Haeffner, and C. W. W?chtler, Observation of Synchronization between Two Quantum van der Pol Oscillators in Trapped Ions,
Phys. Rev. X 16, 021062 (2026).
\bibitem{lab36}Z. Leghtas, S. Touzard, I. M. Pop, A. Kou, B. Vlastakis, A. Petrenko, K. M. Sliwa, A. Narla, S. Shankar, M. J. Hatridge, M. Reagor, L. Frunzio, R. J. Schoelkopf, M. Mirrahimi, and M. H. Devoret, Confining the state of light to a quantum manifold by engineered two-photon loss,
Science 347, 853 (2015).
\bibitem{lab37}K. M. Sliwa, M. Hatridge, A. Narla, S. Shankar, L. Frunzio, R. J. Schoelkopf, and M. H. Devoret,
Reconfigurable Josephson Circulator/Directional Amplifier,
Phys. Rev. X 5, 041020 (2015).
\bibitem{lab38}S. Walter, A. Nunnenkamp, and C. Bruder,
Quantum synchronization of two Van der Pol oscillators,
Ann. Phys. (Berl.) 527, 131 (2015).
\bibitem{lab39}N. R. Bernier, L. D. T\'{o}th, A. Koottandavida, M. A. Ioannou, D. Malz, A. Nunnenkamp, A. K. Feofanov, T. J. Kippenberg,
Nonreciprocal reconfigurable microwave optomechanical circuit,
Nat. Commun. 8, 604 (2017).
\bibitem{lab40}G. A. Peterson, F. Lecocq, K. Cicak, R. W. Simmonds, J. Aumentado, and J. D. Teufel,
Demonstration of Efficient Nonreciprocity in a Microwave Optomechanical Circuit,
Phys. Rev. X 7, 031001 (2017).



\end{thebibliography}
\end{document}